\documentclass[aps,prl,twocolumn,amsmath,amssymb,nofootinbib,superscriptaddress,floatfix]{revtex4-1}

\usepackage{amsmath}
\usepackage{amssymb}
\usepackage{amsthm}
\usepackage[dvipdf]{color}
\usepackage{graphicx}
\usepackage{dcolumn}
\usepackage{bm}

\begin{document}
\title{Topological edge floppy modes in disordered fiber networks}

  \author{Di Zhou}
  \author{Leyou Zhang}
  \author{Xiaoming Mao}

 \affiliation{
 Department of Physics,
  University of Michigan, Ann Arbor, 
 MI 48109-1040, USA
 }

\begin{abstract}
Disordered fiber networks are ubiquitous in a broad range of natural (e.g., cytoskeleton) and manmade (e.g., aerogels) materials. In this paper, we discuss the emergence of topological floppy edge modes in these fiber networks as a result of deformation or active driving. It is known that a network of straight fibers exhibits bulk floppy modes which only bend the fibers without stretching them. We find that, interestingly, with a perturbation in geometry, these bulk modes evolve into edge modes. We introduce a topological index for these edge modes and discuss their implications in biology.
\end{abstract}

\maketitle

\noindent \emph{Introduction --}
Recent theoretical advances in applying concepts of topological states of matter to mechanical systems has led to the burgeoning new field of ``topological mechanics'', where nontrivial topologies of the phonon bands give rise to exotic mechanical and acoustic properties~\cite{Prodan2009,Kane2014,Lubensky2015,Wang2015,Nash2015,Suesstrunk2015,Mousavi2015,Yang2015,
Peano2015,Strohm2005,Sheng2006,Pal2016,He2016,Suesstrunk2016,
Xiao2015a,Rocklin2016,Rocklin2017,Paulose2015a,Paulose2015,Chen2014}.  

\begin{figure}[h]%%%%%%%%%%%%%%
\centering
\includegraphics[width=0.49\textwidth]{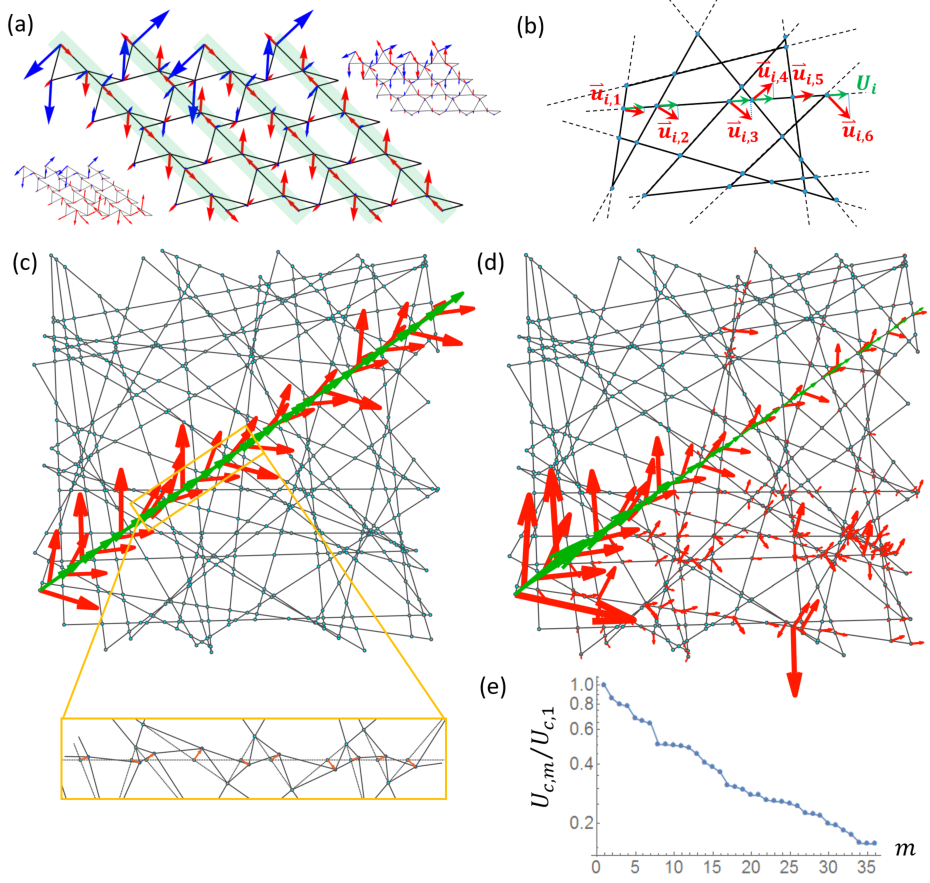}
\caption{(a) A deformed kagome lattice in its critical state (middle, large) between two phases with different topologies in their phonon bands (left and right, small).  These states are related by a soft strain of the lattice that only change the bond angles.  
  Blue and red arrows show a pair of floppy modes, under periodic boundary condition in the horizontal ($x$) direction and open boundary condition in the $y$ direction.  The pair of floppy modes are on the top and bottom edges respectively in the topologically trivial phase (left).  The red mode becomes a bulk mode at the transition (middle, where the cyan stripes show the straight lines of bonds) and shift to the top edge in the topological phase (right).  (b) An example original Mikado network, showing one bulk floppy mode along fiber $i$ (red arrows).  This floppy mode is characterized by a constant longitudinal projection of displacements along the fiber $U_i$ (green arrows), and the displacement vectors of the crosslinkes $\vec{u}_{i,m}$ (red arrows) are perpendicular to the crossing fiber so they are only stretched to second order.  Dangling ends are shown as dashed lines and are ignored in the analysis. (c) Example original Mikado network, showing the bulk floppy mode on the central fiber which is used to obtain the modified Mikado model (red and green arrows showing $\vec{u}_{c,m}^{(0)}$ and $U_{c}^{(0)}$ respectively, magnified by 50 times).  The zoomed in figure below shows details of the displacements ($\vec{u}_{c,m}^{(0)}$ magnified by 10 times) of the central fiber in a local area [boxed in (a)] that leads to the modified Mikado model.  (d) Floppy mode localized on the tail of the central fiber in the  modified Mikado model ($\vec{u}_{c,m}^{(0)}$ too small to be visible). (e) Projection of the floppy mode to each segment $U_{c,m}$ [green arrows in (d)] exponentially decrease from tail ($m=1$) to head ($m=n_c$) on the central fiber. 
}
\label{FIG:fig1}
\end{figure}

Among many different types of topological mechanical systems, a particularly interesting class consists of ``Maxwell lattices'', which are central-force lattices with average coordination number $\langle z \rangle =2d$ where $d$ is the spatial dimension, and are thus at the verge of mechanical instability~\cite{Kane2014,Lubensky2015,Rocklin2016,Rocklin2017,Paulose2015a,Paulose2015,Chen2014}.  
Maxwell lattices host topologically protected phonon edge modes at zero frequency (floppy modes).  These edge modes are governed by the topology of the equilibrium and compatibility matrices of the lattice in the first Brillouin zone, which in turn, are governed by the lattice geometry~\cite{Kane2014}.  A simple two-dimensional example of Maxwell lattice, the deformed kagome lattice, as shown in Fig.~\ref{FIG:fig1}, exhibit different phases where the topological structure changes and the floppy modes localize at different edges~\cite{Rocklin2017}.  In particular, what drives the topological transition here is a soft strain that changes the lattice geometry, where all bonds remain the same length and only the bond angles alter.  At the topological transition, bonds form straight lines and floppy modes penetrate infinitely deep into the bulk, whereas in the two phases below and above the transition, the floppy modes localize at different edges.  In the topologically nontrivial phase all floppy modes localize on the top edge leaving the bottom edge rigid.  
This physics of the Maxwell lattices make them both an interesting topic for theoretical study~\cite{Souslov2009,Mao2010,Ellenbroek2011,Mao2011a,Sun2012,Zhang2015a,Mao2015} and good candidates for the design of novel mechanical metamaterials where the edges can change stiffness by orders of magnitude reversibly~\cite{Rocklin2017}.

Most existing studies of topological mechanics are based on periodic lattices, with only few exceptions~\cite{Sussman2016,Mitchell2016}.  In general, topological order is robust against disorder, because topological attributes are integer valued and remain invariant upon the addition of disorder until they jump to a different integer value.  This robustness has been demonstrated in various periodic lattice systems with weak disorder.  It is thus an intriguing question to ask: can topological edge floppy modes exist in disordered systems that are completely off-lattice?

In this paper, we study floppy edge modes in disordered fiber networks which are not periodic in space (Fig.~\ref{FIG:fig1}b-d).  Fiber networks are ubiquitous in nature, taking the form of cell cytoskeleton and extra-cellular matrix, and in manmade materials, taking the form of fiber hydrogels and aerogels, felt, etc., and exhibit fascinating physics~\cite{Head2003,Wilhelm2003,Gardel2004,Storm2005,Heussinger2006,Broedersz2011,Mao2013b,Mao2013c,Broedersz2014,Sharma2016,Feng2015,Feng2016}.  Using both analytic theory and numerical simulation, we show that topological floppy edge modes exist in these disordered fiber networks, and their existence lead to strongly asymmetric mechanical properties at opposite ends of the fiber network.  These topological edge modes may have interesting consequences in a wide range of problems,  such as cell cytoskeleton under active driving and the design of smart fiber materials.

\noindent \emph{Model and Results --}
We choose the ``Mikado model'', which is a completely off lattice fiber network model~\cite{Head2003,Wilhelm2003}, and modify it for our study of topological edge modes.  The original Mikado model consist of straight fibers randomly placed on a two-dimensional plane, with all crossing points being free hinges (Fig.~\ref{FIG:fig1}b).  The Hamiltonian of a Mikado model can be written as
\begin{align}\label{Hamiltonian}
	H = &\sum_{i=1}^{N_\textrm{fiber}} \sum_{m=1}^{n_{i}-1} 
		\frac{k_{i,m}}{2}\left( |\vec{R}_{i,m}-\vec{R}_{i,m+1}| - {\ell}_{i,m}\right)^2 \nonumber\\
	&+ \sum_{i=1}^{N_\textrm{fiber}} \sum_{m=2}^{n_{i}-1} 
		\frac{\kappa_{i,m}}{2} \left( \Delta \theta_{i,m} \right)^2 ,
\end{align}
where there are $N_\textrm{fiber}$ fibers labeled by $i$, each has $n_i$ crosslinks labeled by $m$, and $\vec{R}_{i,m}$ is the (displaced) position of the $m$-th crosslink on the $i$-th fiber.  The first term denotes central force stretching energy of each fiber segment (bond) between neighboring crosslinks (sites) $m,m+1$ along each fiber $i$, with stretching spring constant $k_{i,m}$ and rest length ${\ell}_{i,m}$.  The second term denotes bending energy of the fiber and $\Delta \theta_{i,m} = \theta_{i,m} - \theta_{i,m-1}$ is the angle change between the two segments meeting at crosslink $m$ along fiber $i$ (here $\theta_{i,m}$ denotes the orientation of the $m$-th segment on fiber $i$, and $\Delta \theta_{i,m}=0$ if fiber $i$ is straight) with bending spring constant $\kappa_{i,m}$.  

In typical fiber networks composed of long slender filaments, the bending stiffness is much smaller compared to the stretching stiffness [$\kappa /(k \ell_0^2) \ll 1$ where $\ell_0$ is the characteristic mesh size, see discussion in the Supplementary Information (SI)].  For our discussion of the topological mechanics we first ignore bending stiffness and treat all fiber segments as central-force springs ($\kappa_{i,m}=0$).  Later we use numerical simulations to verify that the essential conclusion of the asymmetric mechanical properties due to topological edge modes still holds in presence of small bending stiffness.

The original Mikado network display an interesting property: all floppy modes (i.e., modes that do not stretch or compress any bonds) are \emph{bulk modes}.  This can be seen by first apply Maxwell counting to a Mikado network.  The total number of crosslinks is $N_s=\sum_{i=1}^{N_\textrm{fiber}}n_{i}/2$ (remember each crosslink is shared by two fibers) and the total number of bonds is $N_c=\sum_{i=1}^{N_\textrm{fiber}}(n_{i}-1)$ (dangling ends are removed since they don't contribute to mechanical stability).  The number of zero modes is thus equal to the number of fibers $N_0=N_s d-N_c = N_\textrm{fiber}$.  A straightforward decomposition of the $N_\textrm{fiber}$ zero modes is that each fiber carries one zero mode corresponding to the longitudinal displacement of that fiber, while keeping all other fibers intact (the fiber segments crossing the displaced fiber is stretched only to second order of the mode), as shown in Fig.~\ref{FIG:fig1}d~\cite{Heussinger2006}.  It is worth pointing out that these modes are independent but not orthogonal to one another, and they contain the rigid translations and rotation of the whole network.

The original Mikado network can be seen as a disordered analog of the  critical state of the deformed kagome lattice that lies between the topologically trivial and nontrivial phase, in the sense that they both have straight filaments which carry bulk floppy modes (Fig.~\ref{FIG:fig1}a and b).  The deformed kagome lattice exhibit states (related by a soft strain from the critical state) with different topologies where the floppy modes localize at different edges.  Can the Mikado network also exhibit such topological transitions?  The answer is yes.

Because what drives the topological transition and the localization of the floppy modes in the deformed kagome lattice is the change of lattice geometry (in this case induced by the soft strain equivalent to the $\vec{q}=0$ bulk floppy mode), it is natural to consider following bulk floppy modes the original Mikado model and examine their effect on mode localization.  As shown in Fig.~\ref{FIG:fig1}c, we perturb the Mikado model to create a new ground state as follows: one arbitrarily chosen ``central fiber'', $c$, is longitudinally displaced by a small amount $U_{c}^{(0)}$ (each crosslink on this fiber displace by $\vec{u}_{c,m}^{(0)}=U_c^{(0)}(\frac{\sin(\theta_c+\Theta_{c,m})}{\sin\Theta_{c,m}},-\frac{\cos(\theta_c+\Theta_{c,m})}{\sin\Theta_{c,m}})$ where $\theta_c$ is the angle of the central fiber, and $\Theta_{c,m}$ is the intersecting angle between the crossing fiber at crosslink $m$ and the central fiber) following one floppy mode of the original Mikado model.  We choose the convention that if the fiber is pulled in the direction pointing from crosslink 1 to $n_c$ on the central fiber (so crosslink $n_c$ is the ``head'' of motion), $U_{c}^{(0)}>0$, and vice versa, and we ignore the resulting stress (which is second order in $U_{c}^{(0)}$).  This geometric perturbation leads us to a new model which we name ``modified Mikado model''.

\begin{figure}[h]%%%%%%%%%%%%%%
\centering
\includegraphics[width=0.49\textwidth]{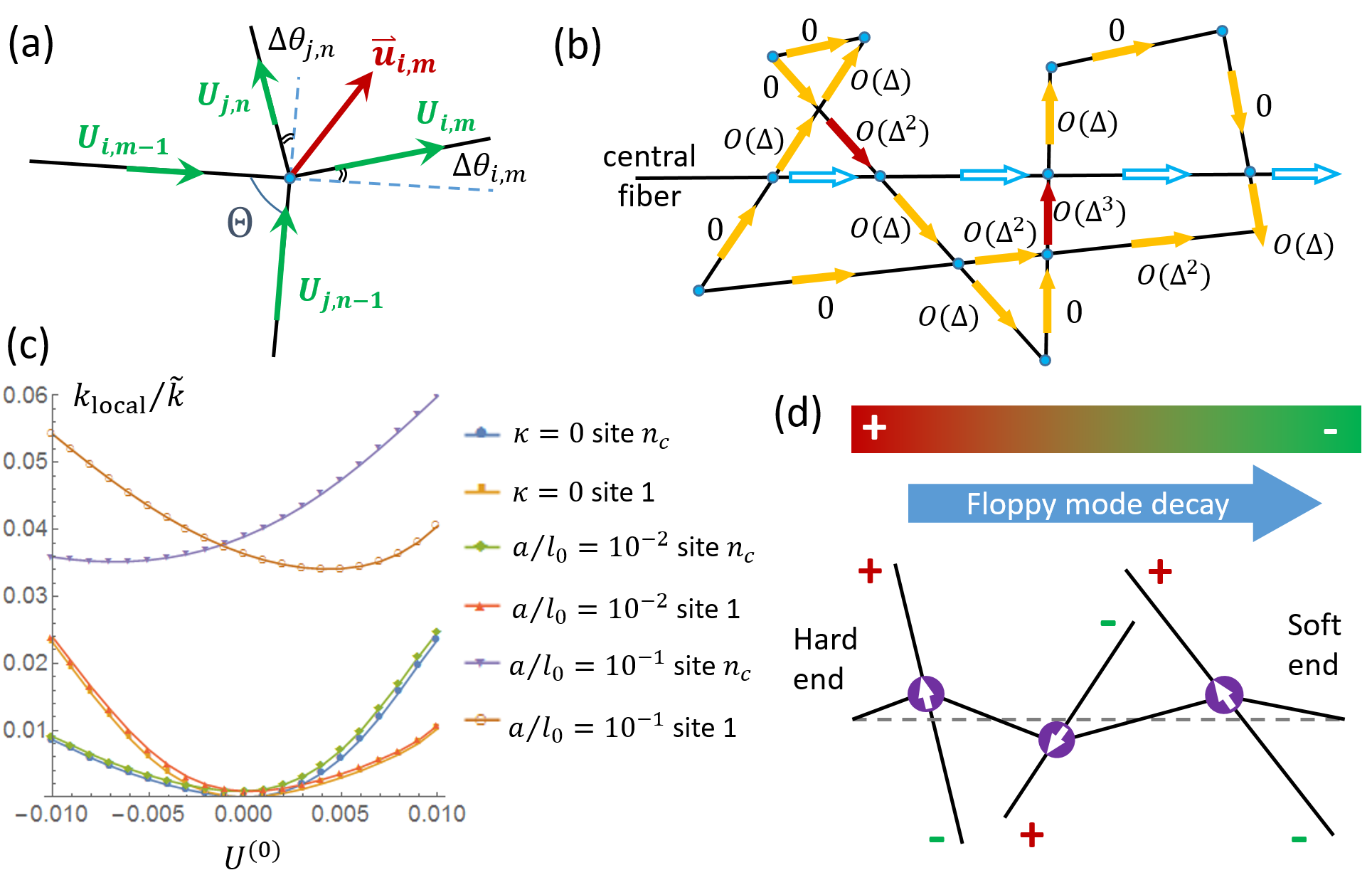}
\caption{(a) Illustration of the transfer matrix [Eq.~\eqref{EQ:TM}] applying on a crosslink.  (b) Displacements propagation (along arrows) and order of magnitude when applying the transfer matrix on the network with boundary condition that only crosslink $1$ of the central fiber has input $U$ (large blue arrows for $O(1)$, smaller arrows for higher order in $\Delta$ and red denotes flow back to the central fiber). (c) Asymmetric edge stiffness at two ends of the central fiber. We perform numerical simulations to measure local stiffness $k_{\textrm{local}}$ against point force on two ends of the central fiber, in modified Mikado models with different $U^{(0)}$.  We show results for both networks with no bending stiffness $\kappa=0$ and with bending stiffness (controlled by fiber thickness $a$ in unit of characteristic mesh size $\ell_0$, and we normalize $k_{\textrm{local}}$ using characteristic spring constant of one segment $\tilde{k}$).  For more details see the SI.  In all cases, the head is significantly more stiff than the tail.
(d) Mikado network under active driving from active crosslinks (marked with arrows) on the central fiber.  The direction of driving is determined by the chirality of the crossing fibers, such that the motors actively move to the ``+'' end.  If all crossing fibers have correlated chirality such that their ``+'' ends are on the left, from Eq.~\eqref{EQ:CFU}, we find that 
the floppy mode on the central fiber exponentially localizes to the left. }
\label{FIG:fig2}
\end{figure}

We then study mechanical properties of the modified Mikado model using both analytical and numerical calculations.  We find that the ``tail'' of the central fiber (i.e., the opposite end to the direction of pulling) host an exponentially localized floppy mode in the modified Mikado model (Fig.~\ref{FIG:fig1}de).  As a result, the local stiffness against a point force is significantly smaller at the tail compared to that at the head of the central fiber (Fig.~\ref{FIG:fig2}c, more details in the SI).

The analytic method we adopt to study the modified Mikado network is based on a transfer matrix that propagate floppy modes through crosslinks in the network.  When the fiber is not straight, instead of a constant longitudinal displacement for the floppy modes in the original Mikado model, the floppy-modes longitudinal displacement is different from segment to segment along a fiber in the modified Mikado model.  Thus, a floppy mode can be characterized either by the displacement of each crosslink, $\{ \vec{u}_{i,m}\}$, or the longitudinal projection of the displacements on each fiber segment $\{ U_{i,m}=\vec{u}_{i,m}\cdot \hat{n}_{i,m} = \vec{u}_{i,m+1}\cdot \hat{n}_{i,m}\}$, where $\hat{n}_{i,m}$ is the unit vector along the $m$-th segment (between site $m$ and $m+1$) on fiber $i$.  
As shown in Fig.~\ref{FIG:fig2}a, the two representations are related by four equations, $U_{i,m-1}=\vec{u}_{i,m}\cdot \hat{n}_{i,m-1}$, $U_{i,m}=\vec{u}_{i,m}\cdot \hat{n}_{i,m}$, $U_{j,n-1}=\vec{u}_{i,m}\cdot \hat{n}_{j,n-1}$, and $U_{j,n}=\vec{u}_{i,m}\cdot \hat{n}_{j,n}$ (assuming the crosslink under consideration is both $m$-th on fiber $i$ and $n$-th on fiber $j$).  Eliminating $\vec{u}_{i,m}$ we get
\begin{eqnarray}\label{EQ:TMequation}
M\left(\begin{array}{c}
U_{i,m-1}\\
U_{j,n-1}\\
\end{array}
\right)=\left(
\begin{array}{c}
U_{i,m}\\
U_{j,n}\\
\end{array}
\right)
\end{eqnarray}
with the transfer matrix 
\begin{eqnarray}\label{EQ:TM}
M = \left(\begin{array}{cc}
\frac{\sin(\Theta_{i,m}-\Delta\theta_{i,m})}{\sin\Theta_{i,m}} & \frac{\sin\Delta\theta_{i,m}}{\sin\Theta_{i,m}}\\
-\frac{\sin\Delta\theta_{j,n}}{\sin\Theta_{i,m}} & \frac{\sin(\Theta_{i,m}+\Delta\theta_{j,n})}{\sin\Theta_{i,m}}\\
\end{array}\right)
\end{eqnarray}
where $\Theta_{i,m} \equiv \theta_{j,n-1} - \theta_{i,m-1}$, $\Delta\theta_{i,m}=\theta_{i,m}-\theta_{i,m-1}$, and $\Delta\theta_{j,n}=\theta_{j,n}-\theta_{j,n-1}$.  This equation serves as a ``transfer matrix'' for segment displacements at crosslinks for an arbitrary floppy mode in the modified Mikado model.  For any input of boundary condition in terms of segment displacements on one end of each fiber (remember the total number of zero mode is equal to the number of fibers), we can calculate the floppy mode displacements throughout the whole network.

With this transfer matrix, we can study general floppy modes in the modified Mikado model.  We are particularly interested in what happens to the floppy mode that was a bulk mode on the central fiber in the original Mikado model (Fig.~\ref{FIG:fig1}bc).  To do this, we take the boundary condition that the first segment of every fiber is given to be $U_{i,1}=0$ if $i\ne c$ and $U_{i,1}=U$ if $i = c$, i.e., only the central fiber has a displacement input along segment 1 (which can be either the head or the tail of the central fiber depending on the pulling that defines the modified Mikado network ground state), while all other fibers are hold fixed at their segment 1.  We then use the transfer matrix [Eq.~\eqref{EQ:TMequation}] to calculate the floppy displacement on the rest of the network.  Figure~\ref{FIG:fig1}c show an example of such exact calculation, where the resulting floppy mode is no longer a bulk mode but instead localizes at the tail of the central fiber.  

To characterize such floppy mode localization we take the following perturbative expansion.  
Because  fibers in the modified Mikado model are close to straight ($U_{c}^{(0)}$ is small), all $\Delta \theta_{i,m}$ are small, which permits a perturbative expansion of the transfer matrix at small bending angles (represented generally by $\Delta$) and allows further analysis.  Following the central fiber, we find that at each crosslink (for more details see the SI),
\begin{eqnarray}
	U_{c,m} = \lbrack 1- \Delta\theta_{c,m} \cot \Theta_{c,m} +\mathcal{O} (\Delta\theta_{c,m}^2)\rbrack
 U_{c,m-1}
\label{EQ:CFU}
\end{eqnarray}
where $\Theta_{c,m}$ is the angle between the central fiber and the crossing fiber at crosslink $m$, and we have used the fact that the input $U_{j,n-1}$ from the fiber which crosses the central fiber is either $0$ (from boundary condition), or of $\mathcal{O}(\Delta^2)$ or higher (from other crosslinks on the central fiber itself through a loop), as shown in Fig.~\ref{FIG:fig2}b.  Such higher order displacements are visible in Fig.~\ref{FIG:fig1}c where we used the full transfer matrix [Eq.~\eqref{EQ:TM}].  Note that this small $\Delta$ expansion also requires that the crossing angles $\Theta_{c,m}$ are not too small (so $\cot \Theta_{c,m}$ does not diverge), a condition naturally satisfied in most fiber networks from excluded volume repulsion.
%A calculation using the exact transfer matrix [Eq.~\eqref{EQ:TM}] is shown in Fig.~\ref{FIG:fig2}c, where higher order displacements on other fibers are visible (they are especially large when fibers cross at small angles).  
%As a result to leading order in $\Delta$ we only need to track the transfer on the central fiber itself

%In order to learn the evolution of $U_{c,m}$ along the central fiber which characterize the localization of the floppy mode, we just need to focus on Eq.~\eqref{EQ:CFU}.  
Equation~\eqref{EQ:CFU} governs the growth and decay of the floppy mode along the central fiber.  If $\cot \Theta>0$, we have $U_{c,m}>(<) U_{c,m-1}$ if $\Delta\theta_{i,m} <(>) 0$ [corresponding to the central fiber bending up (down) at this crosslink], and vice versa (see SI for examples of the geometry).  This is a very general geometric rule for edge floppy modes, which applies to the case of topological kagome lattices as well (e.g., following the two families of vertical lines up in Fig.~\ref{FIG:fig1}c one finds that $U$ increase on both).  This rule can also be used to design new ordered or disordered structures which exhibit tailored distribution of floppy modes (see example in SI).

Now with the general rule of floppy mode evolution at each crosslink, we come back to the question of where the floppy mode localizes in the modified Mikado model.  It is straightforward to see that individually at each crosslink (holding all other crosslinks fixed) the displacement $U_{c,m}^{(0)}$ points to the direction of floppy mode $U_{c,m}$ decrease along the central fiber if $U_{c}^{(0)}>0$ (central fiber pulled towards crosslink $n_c$), and vice versa.  However, we need to rigorously prove that in the modified Mikado model where all crosslinks are displaced along the central fiber at the same time, the disorder averaged (denoted by $\langle \ldots \rangle$) growth rate of the floppy mode
\begin{align}
	\langle \lambda \rangle \equiv 1- \left \langle \frac{U_{c,m+1}}{U_{c,m}} \right\rangle
%= -\left\lbrack\frac{1}{n_c -2}  \sum_{m=2}^{n_c-1}\Delta\theta_{i,m} \cot \Theta_{c,m} \right\rbrack
\end{align}
is positive when $U_{c}^{(0)}>0$ and negative when $U_{c}^{(0)}<0$ (floppy mode localizes on tail), given the condition that different fibers have uncorrelated orientations.  The proof is included in the SI.

The analytic theory discussed above is at zero bending stiffness, but our numerical results  show that when bending stiffness is introduced, the asymmetric stiffness is still significant (Fig.~\ref{FIG:fig2}c).

The floppy edge modes we find in these disordered fiber networks are of the same geometric origin as topological edge floppy modes in periodic lattices.  In discussions above we constructed a real space transfer matrix method that shows the exponential localization of floppy modes on individual fibers.  Next we show that a topological invariant, a generalization of the ``topological polarization'' defined in Ref.~\cite{Kane2014} to disordered networks, can be defined on the central fiber that dictates its edge floppy mode.  In order to do this we start by introducing the compatibility matrix $C_{\beta m}$ which maps site displacements (projected to bond $m$) $U_{c,m}$ onto bond extension $\delta l_{c,\beta}$
\begin{align}\label{EQ:CU}
	\delta l_{c,\beta}=\sum_{m=1}^{n_c}C_{\beta m}U_{c,m}.
\end{align}
The form of $C_{\beta m}$ is determined by the transfer matrix, as detailed in the SI.  We then rewrite this equation in momentum space, where the compatibility matrix takes the form $\tilde{C}(q_1,q_2)$.  Note that it depends on two momenta as a result of disorder (absence of translational invariance) instead of one in the periodic lattice case.  Existence of floppy modes is determined by the equation $\det \tilde{C}=0$ which generally has no solution under periodic boundary condition.  Edge floppy modes under open boundary condition is captured by introducing an extra complex component to the momenta, $k=k'+ik''$.  The sign of $k''$, which governs which end of the fiber the floppy mode localizes to, is determined by a topological invariant, the winding number
\begin{eqnarray}
\mathcal{N}_c =  \frac{1}{n_c}\frac{1}{2\pi {\rm i}}\oint_0^{2\pi}dk\frac{d}{dk}\,{\rm Im}\,\ln\det \tilde{C}(q_1+k,q_2+k) , \quad\quad
\end{eqnarray}
such that $\mathcal{N}_c=0,1$ correspond to floppy mode on the right and left respectively.  The actual solution $k''$ is directly related to the decay rate $\lambda$ on the fiber.  
An expanded discussion of $\mathcal{N}_c$ is in the SI.

\noindent \emph{Discussions --}
In this paper we show that in disordered fiber networks, when individual fibers are pulled, a topological edge floppy mode localizes on the tail of the fiber.  In this section we generalize this conclusion and discuss possible application to experimental systems.

First, the scenario of pulling a fiber in a network occurs broadly in various situations.  For example, in cell-cell and cell-extracellular matrix interactions, actin filaments can exert active pulling on the network, leading to the geometry of fibers being pulled following the network floppy modes, and thus asymmetric stiffness arises, as we discuss above.   The effect that the site of pulling (head) becomes stiff and the opposite end of the fiber (tail) becomes soft, may have interesting consequences in cell mechanics.  
%This may be a mechanism similar to strain stiffening of fiber networks against homogeneous strain, which protects the network from deforming too much at the site of pulling.  
Although the above discussion specialize to the case of one single fiber being pulled, in the SI, we include numerical results for networks in which multiple fibers are pulled simultaneously, where we show edge floppy modes on each pulled fiber.  Moreover, in the modified Mikado network we ignored the (higher order) stress generated in the ground state.  Adding back these residual stresses only shifts the equilibrium position of the head and the tail of the fibers, and the asymmetric stiffness we discuss here remains true (see SI for more discussion).  

Second, although our discussion is based on the simple geometric perturbation that one central fiber is pulled, the transfer matrix method we develop actually applies to more general situation of geometric perturbation of the fiber network, because the exponential increase/decrease of the floppy mode only depend on the relation between the crossing fiber orientation and the direction of the bending of the central fiber.  This type of change of geometry in fiber networks can occur in a rich variety of systems.  For example, in a network where some or all of the crosslinks are active motors which walk on particular directions on the fibers~\cite{Alberts2008,Brangwynne2008,Joanny2009}, such coherent change in geometry can also happen.  As shown in Fig.~\ref{FIG:fig2}d, where a central fiber is crosslinked to other fibers via active motors, and the chirality of the crossing fibers are correlated, a topological edge floppy modes emerge on the central fiber due to the active driving.

\noindent \emph{Acknowledgments --}
This work was supported by the National Science Foundation Grant No. NSF DMR-1609051.

\appendix

\section{Appendix I: Small angle approximation and the disorder averaged decay rate of the edge floppy mode}\label{SEC:SADecay}
In this SI section we start from the floppy mode transfer matrix, Eq.(3) in the main text, discuss its approximations at small $U_c^{(0)}$, and the resulting decay rate of the floppy mode.

For small $U_c^{(0)}$ the new ground state is very close to the original Mikado model with straight fibers.  The only differences are the small bending angles when the central fiber meets the crossing fibers.  When the transfer matrix is applied on these crosslinks, one can expand to first order in the small bending angles (generally denoted as $\Delta$), which is equivalent to first order in $U_c^{(0)}$, and find
%that the central fiber has bending angles $\Delta\theta_{c,m}\ll 1$ at each crosslink, and the crossing fibers 
%The geometric deformation $U_c^{(0)}$ on the head of the fiber in the Mikado network is supposed to be small compared to the bond lengths: $U_c^{(0)}\ll l_{i,m}$. In the new ground state, the bending angles $\Delta\theta_{i,m}$ between the neighboring bonds can be treated perturbatively: $\Delta\theta_{i,m}\ll 1$, which allows us to expand the transfer matrix in orders of the bending angles. Up to the first order of the bending angles, the transfer matrix $M$ in Eq.(\ref{3}) can be simplified as 
\begin{eqnarray}\label{EQ:Msmall}
M & = & \left(\begin{array}{cc}
1-\Delta\theta_{i,m}\cot\Theta_{i,m} & \Delta\theta_{i,m}\csc\Theta_{i,m}\\
-\Delta\theta_{j,n}\csc\Theta_{i,m} & 1+\Delta\theta_{j,n}\cot\Theta_{i,m}\\
\end{array}\right).
\end{eqnarray}
We then use this asymptotic transfer matrix to study the evolution of the floppy mode along the central fiber, with the boundary condition described in the main text, that a displacement $U$ is input on site 1 on the central fiber while site 
$1$ on all other fibers are fixed.  It is easy to see that to $O(\Delta)$, the floppy mode along the central fiber evolve as 
\begin{align}\label{EQ:Um}
	U_{c,m} = (1-\Delta \theta_{c,m}\cot \Theta_{c,m}) U_{c,m-1}
\end{align}
which is Eq.~(4) in the main text.

Next we discuss the disorder average of the mode growth/decay.  As discussed in the main text, when the pulling affect each crosslink $m$ individually (take all other $\vec{u}_{c,n\ne m}=\vec{u}_{i,n}=0$), the factor $1-\Delta \theta \cot \Theta_{c,m}$ has the sign such  that the mode decays (grows) when $U_{c}^{(0)}>0 (<0)$, corresponding to floppy mode localizing at the tail of the pulled central fiber.  Here we discuss the full expression for the disorder averaged decay rate (where all crosslinks displace at the same time)
\begin{align}\label{EQ:Lambda}
	\lambda \equiv 1- \left\langle \frac{U_{c,m+1}}{U_{c,m}} \right\rangle
\end{align}
where $\langle \cdot \rangle $ denote disorder average.  Using the fact that the modified Mikado model ground state is obtained from pulling the central fiber from the original Mikado model, we have (from equation of $\vec{u}_{c,m}^{(0)}$ in main text)
\begin{widetext}
\begin{eqnarray}\label{EQ:Theta}
	\Delta\theta_{c,m} = \frac{U_c^{(0)}}{\ell_{c,m}}\left(\cot\Theta_{c,m}-\cot\Theta_{c,m+1}\right)- \frac{U_c^{(0)}}{\ell_{c,m-1}}\left(\cot\Theta_{c,m-1}-\cot\Theta_{c,m}\right) ,
\end{eqnarray}
to leading order in $U_c^{(0)}$, where $\ell_{c,m}$ is the length of the segment $m$ on the central fiber.  Plug this into the mode evolution [Eq.~\eqref{EQ:Um}], we have the decay rate
\begin{align}
	\lambda = \left\langle \Big\lbrack \frac{U_c^{(0)}}{\ell_{c,m}}\left(\cot\Theta_{c,m}-\cot\Theta_{c,m+1}\right)
					-\frac{U_c^{(0)}}{\ell_{c,m-1}}\left(\cot\Theta_{c,m-1}-\cot\Theta_{c,m}\right) \Big\rbrack \cot \Theta_{c,m} \right\rangle .
\end{align}
Using the fact that the fibers are randomly placed on the 2D plane with no correlation between different fibers, we have that $l_{c,m}$ and $\Theta_{c,m}$ are independent random variables, thus
\begin{align}
	\lambda = \frac{U_c^{(0)}}{\bar{\ell}} \left\langle - \cot\Theta_{c,m+1}\cot\Theta_{c,m} -\cot\Theta_{c,m-1}\cot\Theta_{c,m}+ 2\cot\Theta_{c,m}^2 \right\rangle ,
\end{align}
where $\bar{\ell}$ is the average mesh size.  Further, because $\Theta_{c,m+1}$ and $\Theta_{c,m}$ are also independent variables, and $\langle \cot \Theta_{c,m} \rangle=0$, we have
\begin{align}
	\lambda = 2 \frac{U_c^{(0)}}{\bar{\ell}} \langle \cot\Theta_{c,m}^2 \rangle .
\end{align}
Therefore $\lambda$ has the same sign as $U_c^{(0)}$, showing that the floppy mode always \emph{localize on the tail of the pulled central fiber}.

It is worth noting that $\langle \cot\Theta_{c,m}^2 \rangle$ is actually divergent, which is an artifact of our small angle approximation.  When the two crossing fibers are too close to parallel ($\Theta_{c,m}\to 0 $ or $\pi$), we have $\sin \Theta_{c,m}\to 0$ and the crosslink displacement diverges, making the small angle $\Delta \theta_{c,m}$ approximation invalid.  In reality such near-parallel crossings are naturally avoided by excluded volume interactions between the fibers so the divergence is regularized.
%Including higher order terms will regularize this divergence.  Such exponential decay of the floppy mode is verified in our numerical simulations (see main text).
\end{widetext}

\section{Appendix II: Crosslink geometry for floppy mode decay or growth}
In this SI section we illustrate geometries that give rise to growth or decay of the floppy mode (FM) as well as using this principle to design new structures that exhibit edge floppy modes.

\begin{figure}[h]%%%%%%%%%%%%%%
\centering
\includegraphics[width=0.45\textwidth]{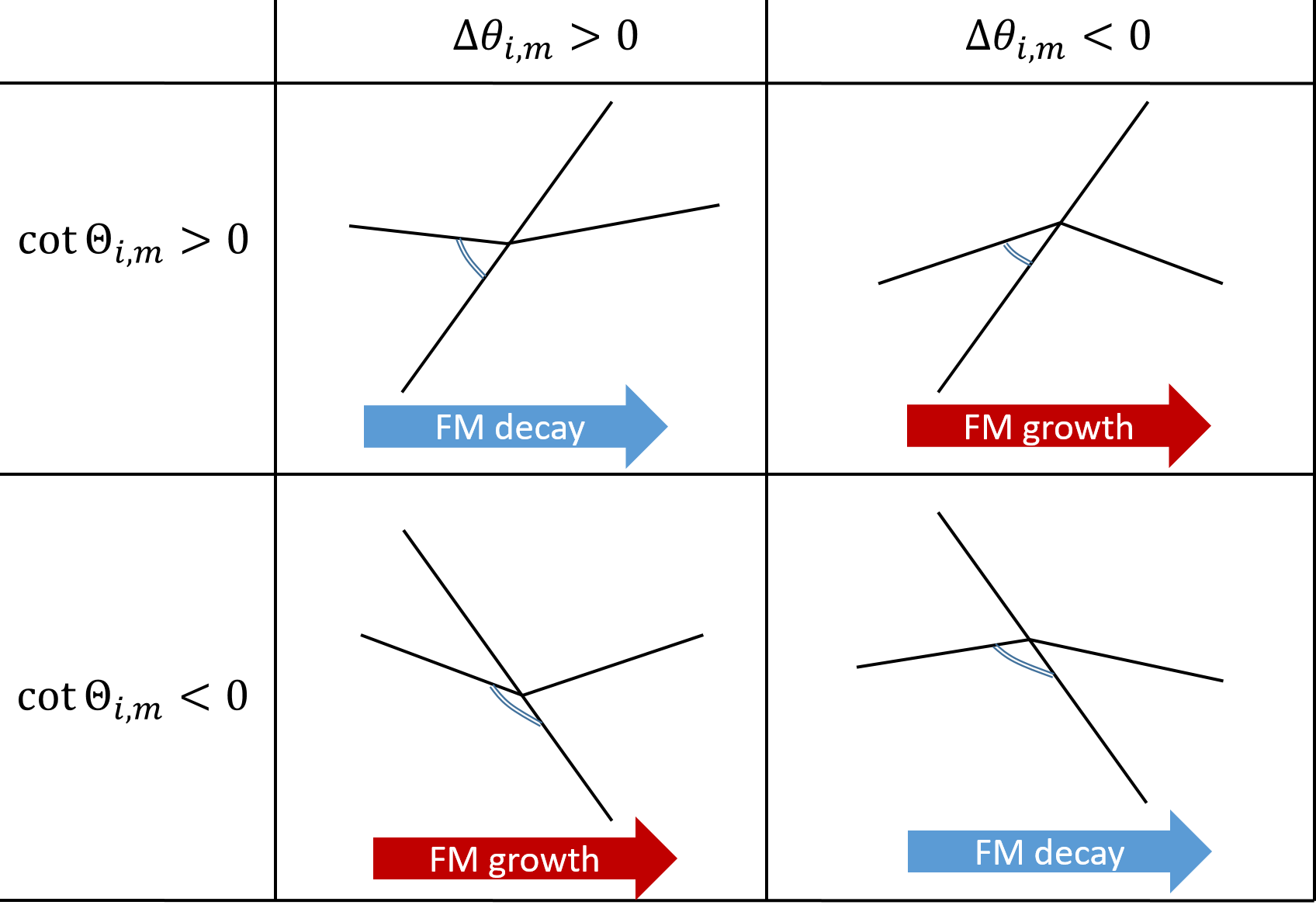}
\caption{4 different geometries of the crosslink and their resulting floppy mode evolution.  Fiber $i$ is the horizontal one and $j$ is the vertical one.  Note that whether the crossing fiber bends or not is irrelevant for the floppy mode on the fiber $i$.}
\label{FIG:geometry}
\end{figure}

In Fig.~\ref{FIG:geometry} we show various crosslink geometries, for different choices of crossing angle (from central fiber to the crossing fiber) $\Theta_{i,m}$ and bending angle of the central fiber at this crosslink $\Delta \theta_{i,m}$.  Because from Eq.~(4) in the main text, that
\begin{eqnarray}
	U_{c,m} = (1- \Delta\theta_{c,m} \cot \Theta_{c,m} +\mathcal{O} (\Delta\theta_{c,m}^2)) U_{c,m-1}.
\label{EQ:CFU}
\end{eqnarray}
The growth or decay of the floppy mode is simply controlled by the combination $\Delta\theta_{c,m} \cot \Theta_{c,m}$.  Thus we have 4 different cases as shown in Fig.~\ref{FIG:geometry}.

\begin{figure}[h]%%%%%%%%%%%%%%
\centering
\includegraphics[width=0.45\textwidth]{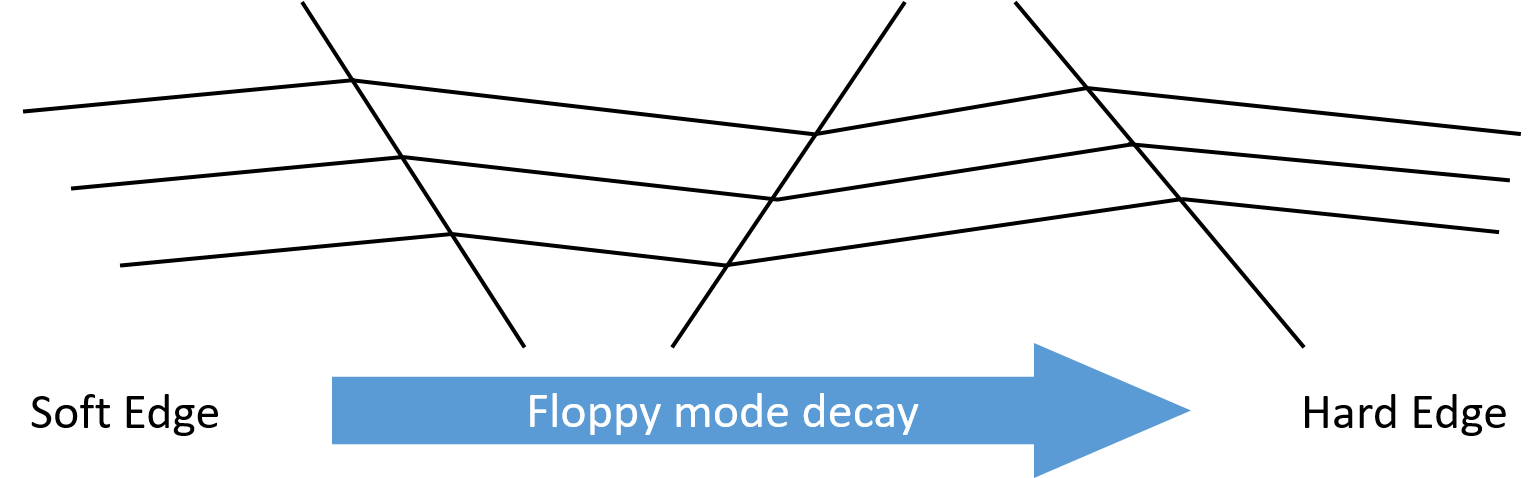}
\caption{A structure where crossing fibers (vertical ones) of alternating $\Theta_{c,m}$ are arranged, such that all horizontal fibers exhibit floppy edge modes localized on their left.}
\label{FIG:design}
\end{figure}

Using this principle we can design new ordered or disordered structures that exhibit floppy modes on chosen edges.  Fig.~\ref{FIG:design} show one such example.

\section{Appendix III: Topological index of the floppy mode in the modified Mikado model}
The robustness of the floppy mode localization at the tail of the central fiber in the modified Mikado model calls for a definition of a topological index.  In this SI section we define this topological index, which extends the ``topological polarization'' concept from Ref.~\cite{Kane2014} to disordered fiber networks.

We start by introducing the compatibility matrix $C_{\beta m}$ that maps the displacement longitudinal projection $U_{c,m}$ to the bond extension $\delta l_{c,\beta}$, 
\begin{align}\label{EQ:CU}
	\delta l_{c,\beta}=\sum_{m=1}^{n_c}C_{\beta m}U_{c,m}
\end{align}
Using the floppy mode we derived in the main text and in the SI Sec.~\ref{SEC:SADecay} (in small angle approximation), we have 
\begin{eqnarray}\label{EQ:CReal}
C_{\beta m}=(1-\Delta\theta_{c,m}\cot\Theta_{c,m})^{-1}\delta_{\beta+1, m}-\delta_{\beta, m} .
\end{eqnarray}
With open boundary condition (OBC) on both ends of the central fiber, $C_{\beta m}$ is a $(n_c-1)\times n_c$ matrix which maps the $n_c$-dimensional $U_c$ space to the $n_c-1$-dimensional $\delta l_c$ space.  Thus its null space at least contains one floppy mode on the central fiber.  In the rest of this section, for simplicity we define $c_{m}\equiv(1-\Delta\theta_{c,m}\cot\Theta_{c,m})^{-1}$. 

Similar to the discussion of in the periodic lattices, we rewrite Eq.~\eqref{EQ:CU} in a ``quasi'' momentum space, where although no periodic lattice structure exist, we Fourier transform based on the crosslink labels along the central fiber.  The Fourier transform and its inverse are defined as
\begin{align}\label{B2}
  \tilde{U}_c(q) &= \sum_{m=1}^{n_c}U_{c,m}e^{-iq m}, \nonumber\\
	U_{c,m} &= \frac{1}{n_c}\sum_{j=1}^{n_c}\tilde{U}_c(q)e^{{\rm i}q m},
\end{align}
where $q=2\pi j/n_c$ with $j$ being integers from $1$ to $n_c$.  In defining this Fourier transform we have assumed periodic boundary conditions (PBC), where the $(n_c+1)$th site is the 1st site, and energy term between them is determined by $\Theta_{c,1}$.  
%and an auxiliary $n_c$ segment was introduced between crosslink $n_c$ and $1$ to form a ring. 
 In this quasi-momentum space, Eq.~\eqref{EQ:CU} takes the form
\begin{align}\label{EQ:CUMome}
	\delta \tilde{l}_c(q_1)=\frac{1}{n_c}\sum_{q_2}\tilde{C}(q_1,q_2)\tilde{U}_c(q_2).
\end{align}
It is worth noting that, unlike periodic lattices, where the Fourier transform of the compatibility matrix reduces to $\delta(q_1-q_2)\tilde{C}(q_1)$ out of translational invariance, here the compatibility matrix still depends on both momenta.  

Nevertheless, floppy modes still correspond to null space of $\tilde{C}(q_1,q_2)$ in momentum space.  In order to have floppy modes, the condition
\begin{align}\label{EQ:detCt}
	\det \tilde{C} =0
\end{align}
has to be met (where the determinant is taken in the $q_1,q_2$ space).  In particular, from Eq.~\eqref{EQ:CReal} we have
\begin{align}
	\tilde{C}(q_1,q_2) = \tilde{c}(q_1-q_2) e^{i q_1}- n_c \delta_{j_1,j_2}
\end{align}
where $q_1=2\pi j_1/n_c, q_2=2\pi j_2/n_c,$
\begin{align}
	\tilde{c}_{q}=\sum_{m=1}^{n_c}e^{-{\rm i}m q }c_m
\end{align}
is the Fourier transform of $c_{m}\equiv(1-\Delta\theta_{c,m}\cot\Theta_{c,m})^{-1}$.

Similar to the periodic lattices, in general the condition~\eqref{EQ:detCt} is not satisfied if $q_1, q_2$ are real numbers (no floppy modes on a ring).  However, if we introduce an imaginary part to the momentum, so that $q \to q+k$ where $k=k' + ik''$ is a complex variable, the equation $\det \tilde{C}=0$ can be solved.  The physical meaning of introducing $k$ is that instead of requiring PBC $U_{c,m+n_c} = U_{c,m}$, now
\begin{align}
	U_{c,m+n_c} = U_{c,m} e^{(ik'-k'') n_c}
\end{align}
so we effectively ``decouple'' the the two ends of chain.  Instead of being a ``ring'' under PBC, it is now a ``spiral'' with the pitch determined by $k$.  With $\tilde{C}(q_1+k,q_2+k) = \tilde{c}(q_1-q_2) e^{i (q_1+k)}- n_c \delta_{j_1,j_2}$ we find that  $\det \tilde{C}=0$ reduces to
\begin{align}\label{EQ:Solvk}
	1-e^{i k n_c} \prod_{m=1}^{n_c} c_m =0 ,
\end{align}
leading to the solution
\begin{align}
	k' &= 0, \nonumber\\
	k'' &=\frac{1}{n_c} \log \left(\prod_{m=1}^{n_c} c_m \right) .
\end{align}
This agrees with the total decay of the floppy mode on the fiber obtained in real space, that
\begin{align}
	U_{c,n_c}/U_{c,1} = \prod_{m=2}^{n_c} c_m^{-1}.
\end{align}

Whether the floppy mode is localized on the left or the right end of the fiber is captured by the sign of $k''$, which is determined by whether the product $\prod_{m=2}^{n_c} c_m$ is less than or greater than 1.  Following the construction of the topological polarization in the regular lattices, this is related to the winding number of the phase of $\det \tilde{C}$ when $k$ goes around the first Brillouin zone $k=0 \to 2\pi$,
\begin{eqnarray}\label{B3}
\mathcal{N}_c =  \frac{1}{n_c}\frac{1}{2\pi {\rm i}}\oint_0^{2\pi}dk\frac{d}{dk}\,{\rm Im}\,\ln\det \tilde{C}(q_1+k,q_2+k) , \quad\quad
\end{eqnarray}
where the factor of $1/n_c$ comes from the fact that the equation~\eqref{EQ:Solvk} is actually $n_c$ degenerate from the $n_c\times n_c$ matrix determinant.
This winding number counts the solution of $k$ inside the unit circle, corresponding to $k''<0$, which is a floppy mode on the right boundary.  Thus, when $\mathcal{N}_c=0$ the floppy mode is localized on the left, and when $\mathcal{N}_c=1$ the floppy mode is localized on the right.

Finally we need to comment on the physical meaning of the topological invariant $\mathcal{N}_c$ in the disordered fiber network model.  The whole formulation from compatibility matrix to the definition of the topological invariant $\mathcal{N}_c$ is rather general: we assume that the decay of the floppy mode from site $m-1$ to $m$ is controlled by the series $\{c_m\}$,
\begin{align}
	U_{c,m} = c_m^{-1} U_{c,m-1} .
\end{align}
When the ground state is generated by pulling the central fiber, as described in the main text, $c_m$ is determined by $c_{m}\equiv(1-\Delta\theta_{c,m}\cot\Theta_{c,m})^{-1}$.

The definition of the compatibility matrix and the topological invariant is independent of the actual form of $c_m$.  As long as we can write down Eq.~\eqref{EQ:CU}, all discussions in this section follows.  Thus, it may seems surprising that for any disordered fiber network, as long as the spectrum is gapped on the central fiber (meaning that $\det \tilde{C}$ has no solution under PBC), the topological invariant $\mathcal{N}_c$ can always be defined.  We have to point out that although this is true (that one always get $\mathcal{N}_c=0$ or $1$), it doesn't automatically mean a well defined localized mode.  The physical meaning of $\mathcal{N}_c$ in this discussion is simply the sign of $k''$ that characterizes the ratio between $U_{c,n_c}$ and $U_{c,1}$.  It doesn't guarantee a coherently decaying or growing floppy mode through the fiber.  The rigorous ``exponential localization'' of the mode requires that $c_m$ consistently $>1$ or $<1$ on most sites.  This impose additional requirements on the details of the disorder.  The example of modified Mikado model, as we discuss in the main text, is indeed characterized by coherent growth/decay along the central fiber, because it has the special geometry of being generated by pulling this fiber along the floppy mode of the straight state.  In the main text we discuss another case, where crosslinks are active and the correlation of the fiber polarizations guarantee the exponential localization.  This type of condition is usually not satisfied on a fiber network with generic disorder.  Thus, in order to have exponentially localized edge floppy mode, we need to show that the decay rate as defined in Eq.~\eqref{EQ:Lambda} has consistent signs throughout the chain.

\section{Appendix IV: Numerical simulation of the modified Mikado model and the addition of bending stiffness}
In this section we discuss how we numerically simulate the modified Mikado model, to characterize (i) floppy modes localized on the tail of the central fiber, and (ii) asymmetric edge stiffness, in presence of bending stiffness, as a result of the edge floppy mode.
\subsection{A. Simulation protocol and calculating the edge floppy mode}
We first generate samples of the original Mikado model by creating $N_{\textrm{fiber}}=50$ straight fibers. The orientations of the fibers, $\theta_i$ are randomly distributed from $0$ to $2\pi$, with the constraints $\frac{\pi}{20}<|\theta_i-\theta_j|<\pi-\frac{\pi}{20}$ or $\pi+\frac{\pi}{20}<|\theta_i-\theta_j|<2\pi-\frac{\pi}{20}$, $\forall i, j =  1, 2, ..., 50$, to mimic excluded volume interactions between the fibers and such that the small angle approximation of the transfer matrix is valid. The centers of the fibers are randomly distributed in an $L\times L$ box, with $L=10$. The fibers are infinitely long to start with, but then we keep only fiber-fiber intersections (crosslinks) within the $L\times L$ box, and remove the dangling ends of the fibers.  The resulting networks look like Fig.~1c in the main text.  The mesh size of the network is characterized by the length scale 
\begin{align}
\ell_0 = L/N_\textrm{fiber},
\end{align}
where $L$ characterize the length scale of the fibers and $N_\textrm{fiber}$ characterize the number of crosslinks on one fiber.  In the networks we generated the measured mesh size is $\bar{\ell} \simeq 0.33$ which differ from $\ell_0=0.2$ by a geometric constant of $O(1)$.  We will later use $\ell_0$, the mesh size length scale, as a natural length unit in presenting the results.

The modified Mikado model is obtained by randomly choosing a central fiber from the original Mikado model, and apply the bulk floppy mode $\vec{u}_{c,m}^{(0)}$ of the central fiber on the original Mikado model, as described in the main text.  

We then calculate the floppy mode on the modified Mikado model using the transfer matrix method we derived in the main text.  Taking $U_{c}^{(0)}=0.01$, and the BC that only a displacement $U_{c,1}>0$ is applied on site 1 of the central fiber while keeping site 1 of all other fibers fixed, we apply the transfer matrix Eq.~(3) from the main text throughout the network to find displacements of all sites (the magnitude of $U_{c,1}$ is not important because we are doing linear analysis).  The result is shown in Fig.~1de.

\subsection{B. Adding bending stiffness and measure asymmetric local stiffness on two ends of the central fiber}
In this subsection we characterize the asymmetric local stiffness in presence of fiber bending stiffness.  In order to provide a realistic mechanical description of a fiber network, instead of assigning uniform $k_{i,m}$ to all segments
and $\kappa_{i,m}$ to all crosslinks, we model the fibers as thin rods of radius $a$ and Young's modulus $Y$.

The continuum mechanics of thin rods is described by the elastic energy
\begin{align}
	H_{\textrm{rod}} = \frac{1}{2} Y \pi a^2 \int_{0}^{\ell} \left(  \frac{du}{ds} \right)^2 ds + \frac{1}{2} \frac{Y \pi a^4}{4} \int_{0}^{\ell} \left(  \frac{d\theta}{ds} \right)^2 ds,
\end{align}
where the first term is the stretching energy  and the second term is the bending energy ($s$ is the coordinate along the arc length and $u(s), \theta(s)$ are the displacement and angle at $s$).

Assuming segments from the fiber network are thin rods described by this equation, we find
\begin{align}\label{EQ:kkappa}
	k_{i,m} &= \frac{Y \pi a^2}{\ell_{i,m}} , \nonumber\\
	\kappa_{i,m} &= \frac{Y \pi a^4}{4} \frac{2}{\ell_{i,m-1}+\ell_{i,m}} .
\end{align}
where $k_{i,m}, \kappa_{i,m}$ are stretching and bending spring constants in the Hamiltonian [Eq.(1) of the main text].  The first equation on stretching spring constant is quite straight forward from the definition of Young's modulus.  The second equation on bending spring constant comes from the harmonic mean of the bending constants of the two fiber segments meeting at site $m$.  Note that although the rod material is homogeneous along the fiber, the bending spring constant [as defined in Eq.(1) in the main text] is a function of the segment length.  The harmonic mean comes from the fact that the two segments are connected in series at site $m$.  A similar construction that discretize continuous bending of a rod onto a lattice model was used in Ref.~\cite{Mao2013b}.  Following the detailed discussion in Ref.~\cite{Mao2013b} one can show that the bending energy of the whole fiber maps to a sum of discrete bending terms $\sum_{m}\frac{\kappa_{i,m}}{2} (\Delta \theta_{i,m})^2$ [as appeared in Eq.(1) in the main text].

\begin{figure}[h]%%%%%%%%%%%%%%
\centering
\includegraphics[width=0.48\textwidth]{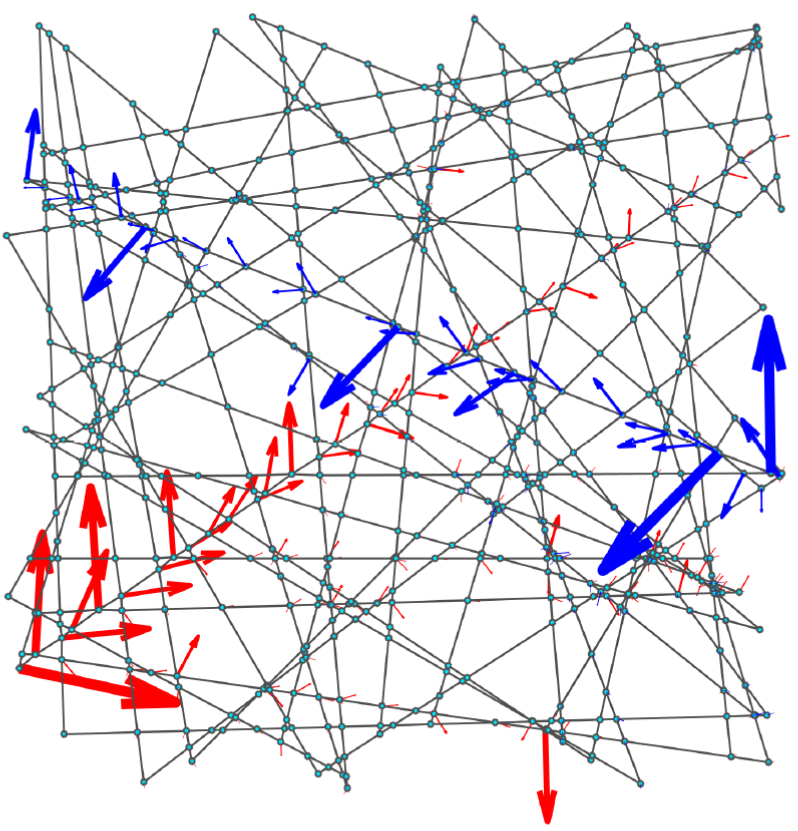}
\caption{A modified Mikado network in which two fibers are pulled.  Blue and red arrows show displacements of the two floppy modes on these two fibers (calculated separately using the boundary condition described in the paper for generating edge floppy mode on these two fibers individually).}
\label{FIG:Multi}
\end{figure}

Using these spring constants we then measure local mechanical response of the modified Mikado model.  The purpose is to characterize the asymmetric local stiffness of the head and the tail of the central fiber, as a result of the exponentially localized floppy mode.  In order to do this, we apply a small force $f_t$ (direction along the central fiber) on the measurement site (either site $1$ or $m$) on the central fiber, leaving the other end of the central fiber free, while fixing both ends of all other fibers.  We equilibrate the network using gradient descent algorithm and measure the displacement of the measurement site projected to the longitudinal direction, $u_t$.  The local stiffness at the measured end of the central fiber is then given by
\begin{align}
	k_{\textrm{local}} = f_t / u_t .
\end{align}
The magnitude of $f_t$ is chosen to be small enough so that the measurement is in the linear elasticity regime (where $f_t$ vs $u_t$ curve is sufficiently straight across positive and negative $f_t$).

We measure the local stiffness $k_{\textrm{local}}$ as a function of both $U_{c}^{(0)}$ and bending stiffness.  We take $U_c^{(0)}=(-10, -9, ..., -1, 0, 1, ..., 9, 10)\times 10^{-3}$ applied on the central fiber along the direction from site $1$ to site $m$.  When $U_{c}^{(0)}>0$ the site $1$ is the tail and the site $m$ is the head, and vice versa.  
%We make measurements at 3 different choices of bending stiffness: (i) central force 

The choice of bending stiffness is based on the following consideration.  When we measure mechanical properties in simulation, the actual control parameter that comes from fiber properties is actually the dimensionless combination
\begin{align}
	\left\langle \frac{\kappa_{i,m}}{k_{i,m}\ell_{i,m}^2} \right\rangle \sim \left( \frac{a}{\ell_0} \right)^2
\end{align}
which is controlled by the ratio between the rod radius and the characteristic mesh size.  The overall factor of the Young's modulus can actually be factorized out in the Hamiltonian.  In practice, we keep the mesh size fixed while vary the fiber radius to obtain different values of this ratio, and express our measurement of stiffness in unit of characteristic  spring constant 
\begin{align}
	\tilde{k} = \frac{Y \pi a^2}{\ell_0} .
\end{align}

Thus, we take 3 different choices of bending stiffness: (i) $a/\ell_0=10^{-2}$ but take $\kappa_{i,m}=0$ at all crosslinks (central force network) and $k_{i,m}$ determined from Eq.~\eqref{EQ:kkappa}, (ii) $a/\ell_0=10^{-2}$, and (iii) $a/\ell_0=10^{-1}$.  For both (ii) and (iii) the spring constants are determined by Eq.~\eqref{EQ:kkappa}.

Following this construction, we generate 10 samples of modified Mikado models, and randomly take 100 fibers from these networks as the central fiber to collect the data for local stiffness.  In addition, in generating these networks we exclude fiber positions which lead to crosslinks too close to one another (distance smaller than $L/200$), to reflect finite size of the crosslinkers.  
Our results are presented in Fig.~2c in the main text.

\begin{widetext}
 \section{Appendix V: Additional results}
\subsection{A. Multiple pulled fiber}
In Fig.~\ref{FIG:Multi} we show that in a modified Mikado network where multiple central fibers are pulled simultaneously, each of these fibers individually host an edge floppy mode.

\subsection{B. Keeping residual stress of pulling}
In Fig.~\ref{FIG:stress} we show the force-displacement curve of a central fiber in a Mikado network where residual stress is kept, i.e., we do not ignore the stress generated by pulling the central fiber to reach the modified Mikado model.  Alternatively speaking, the force $f_t$ is exerted on site $n_c$ on the central fiber ($f_t=0$ correspond to the original Mikado network, while the two ends of all other fibers are fixed), and we record the displacement of this site.  When $f_t>0, u_t>0$ the site $n_c$ is the head and exhibit large local stiffness (slope of the curve), and when $f_t<0, u_t<0$ the site $n_c$ is the tail and exhibit small local stiffness.  Note that the slope at $f_t=u_t=0$ is 0, because we are exiting the bulk floppy mode (in linear regime) of the original Mikado network.
\begin{figure}[h]%%%%%%%%%%%%%%
\centering
\includegraphics[width=0.85\textwidth]{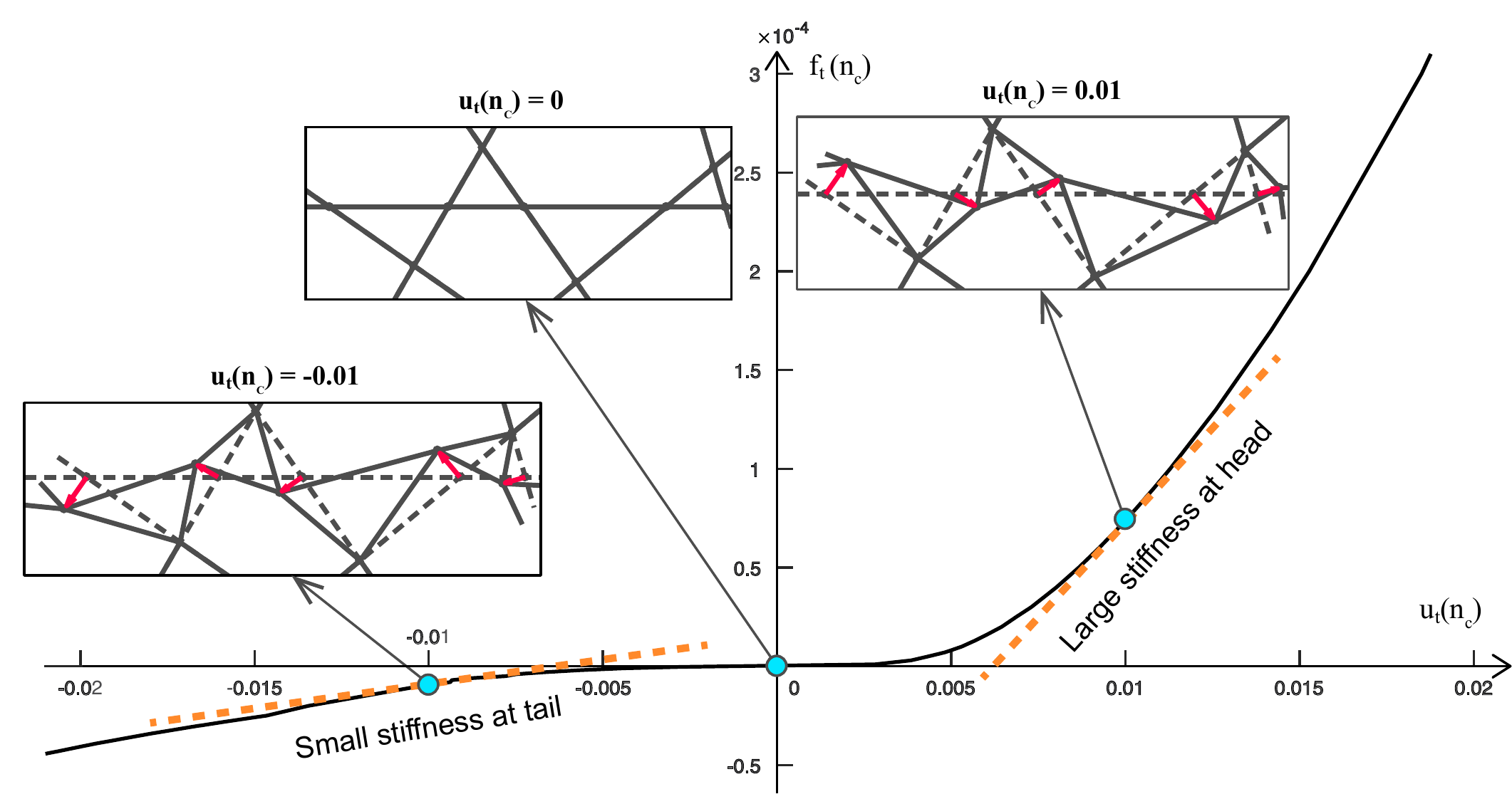}
\caption{Force-displacement curve of exerting force (parallel to fiber) at the end of one fiber in the Mikado network.  $f_t>0$ is pulling the site $n_c$ out and $f_t<0$ is pushing it in.  This corresponds to generate the modified Mikado network while keeping all residual stress in.  The slope of the curve correspond to local stiffness, which is the same as what is measured in Fig.2c in the main text (note that Fig.2c in the main text is the result of disorder average of 100 fibers and the curve in this figure is from 1 randomly chosen fiber).}
\label{FIG:stress}
\end{figure}
\end{widetext}


\begin{thebibliography}{44}%
\makeatletter
\providecommand \@ifxundefined [1]{%
 \@ifx{#1\undefined}
}%
\providecommand \@ifnum [1]{%
 \ifnum #1\expandafter \@firstoftwo
 \else \expandafter \@secondoftwo
 \fi
}%
\providecommand \@ifx [1]{%
 \ifx #1\expandafter \@firstoftwo
 \else \expandafter \@secondoftwo
 \fi
}%
\providecommand \natexlab [1]{#1}%
\providecommand \enquote  [1]{``#1''}%
\providecommand \bibnamefont  [1]{#1}%
\providecommand \bibfnamefont [1]{#1}%
\providecommand \citenamefont [1]{#1}%
\providecommand \href@noop [0]{\@secondoftwo}%
\providecommand \href [0]{\begingroup \@sanitize@url \@href}%
\providecommand \@href[1]{\@@startlink{#1}\@@href}%
\providecommand \@@href[1]{\endgroup#1\@@endlink}%
\providecommand \@sanitize@url [0]{\catcode `\\12\catcode `\$12\catcode
  `\&12\catcode `\#12\catcode `\^12\catcode `\_12\catcode `\%12\relax}%
\providecommand \@@startlink[1]{}%
\providecommand \@@endlink[0]{}%
\providecommand \url  [0]{\begingroup\@sanitize@url \@url }%
\providecommand \@url [1]{\endgroup\@href {#1}{\urlprefix }}%
\providecommand \urlprefix  [0]{URL }%
\providecommand \Eprint [0]{\href }%
\providecommand \doibase [0]{http://dx.doi.org/}%
\providecommand \selectlanguage [0]{\@gobble}%
\providecommand \bibinfo  [0]{\@secondoftwo}%
\providecommand \bibfield  [0]{\@secondoftwo}%
\providecommand \translation [1]{[#1]}%
\providecommand \BibitemOpen [0]{}%
\providecommand \bibitemStop [0]{}%
\providecommand \bibitemNoStop [0]{.\EOS\space}%
\providecommand \EOS [0]{\spacefactor3000\relax}%
\providecommand \BibitemShut  [1]{\csname bibitem#1\endcsname}%
\let\auto@bib@innerbib\@empty
%</preamble>
\bibitem [{\citenamefont {Prodan}\ and\ \citenamefont
  {Prodan}(2009)}]{Prodan2009}%
  \BibitemOpen
  \bibfield  {author} {\bibinfo {author} {\bibfnamefont {E.}~\bibnamefont
  {Prodan}}\ and\ \bibinfo {author} {\bibfnamefont {C.}~\bibnamefont
  {Prodan}},\ }\href {\doibase 10.1103/PhysRevLett.103.248101} {\bibfield
  {journal} {\bibinfo  {journal} {Phys. Rev. Lett.}\ }\textbf {\bibinfo
  {volume} {103}},\ \bibinfo {pages} {248101} (\bibinfo {year}
  {2009})}\BibitemShut {NoStop}%
\bibitem [{\citenamefont {Kane}\ and\ \citenamefont
  {Lubensky}(2014)}]{Kane2014}%
  \BibitemOpen
  \bibfield  {author} {\bibinfo {author} {\bibfnamefont {C.~L.}\ \bibnamefont
  {Kane}}\ and\ \bibinfo {author} {\bibfnamefont {T.~C.}\ \bibnamefont
  {Lubensky}},\ }\href@noop {} {\bibfield  {journal} {\bibinfo  {journal} {Nat.
  Phys.}\ }\textbf {\bibinfo {volume} {10}},\ \bibinfo {pages} {39} (\bibinfo
  {year} {2014})}\BibitemShut {NoStop}%
\bibitem [{\citenamefont {Lubensky}\ \emph {et~al.}(2015)\citenamefont
  {Lubensky}, \citenamefont {Kane}, \citenamefont {Mao}, \citenamefont
  {Souslov},\ and\ \citenamefont {Sun}}]{Lubensky2015}%
  \BibitemOpen
  \bibfield  {author} {\bibinfo {author} {\bibfnamefont {T.~C.}\ \bibnamefont
  {Lubensky}}, \bibinfo {author} {\bibfnamefont {C.}~\bibnamefont {Kane}},
  \bibinfo {author} {\bibfnamefont {X.}~\bibnamefont {Mao}}, \bibinfo {author}
  {\bibfnamefont {A.}~\bibnamefont {Souslov}}, \ and\ \bibinfo {author}
  {\bibfnamefont {K.}~\bibnamefont {Sun}},\ }\href@noop {} {\bibfield
  {journal} {\bibinfo  {journal} {Reports on Progress in Physics}\ }\textbf
  {\bibinfo {volume} {78}},\ \bibinfo {pages} {073901} (\bibinfo {year}
  {2015})}\BibitemShut {NoStop}%
\bibitem [{\citenamefont {Wang}\ \emph {et~al.}(2015)\citenamefont {Wang},
  \citenamefont {Lu},\ and\ \citenamefont {Bertoldi}}]{Wang2015}%
  \BibitemOpen
  \bibfield  {author} {\bibinfo {author} {\bibfnamefont {P.}~\bibnamefont
  {Wang}}, \bibinfo {author} {\bibfnamefont {L.}~\bibnamefont {Lu}}, \ and\
  \bibinfo {author} {\bibfnamefont {K.}~\bibnamefont {Bertoldi}},\ }\href@noop
  {} {\bibfield  {journal} {\bibinfo  {journal} {Physical review letters}\
  }\textbf {\bibinfo {volume} {115}},\ \bibinfo {pages} {104302} (\bibinfo
  {year} {2015})}\BibitemShut {NoStop}%
\bibitem [{\citenamefont {Nash}\ \emph {et~al.}(2015)\citenamefont {Nash},
  \citenamefont {Kleckner}, \citenamefont {Read}, \citenamefont {Vitelli},
  \citenamefont {Turner},\ and\ \citenamefont {Irvine}}]{Nash2015}%
  \BibitemOpen
  \bibfield  {author} {\bibinfo {author} {\bibfnamefont {L.~M.}\ \bibnamefont
  {Nash}}, \bibinfo {author} {\bibfnamefont {D.}~\bibnamefont {Kleckner}},
  \bibinfo {author} {\bibfnamefont {A.}~\bibnamefont {Read}}, \bibinfo {author}
  {\bibfnamefont {V.}~\bibnamefont {Vitelli}}, \bibinfo {author} {\bibfnamefont
  {A.~M.}\ \bibnamefont {Turner}}, \ and\ \bibinfo {author} {\bibfnamefont
  {W.~T.}\ \bibnamefont {Irvine}},\ }\href@noop {} {\bibfield  {journal}
  {\bibinfo  {journal} {Proceedings of the National Academy of Sciences}\
  }\textbf {\bibinfo {volume} {112}},\ \bibinfo {pages} {14495} (\bibinfo
  {year} {2015})}\BibitemShut {NoStop}%
\bibitem [{\citenamefont {S{\"u}sstrunk}\ and\ \citenamefont
  {Huber}(2015)}]{Suesstrunk2015}%
  \BibitemOpen
  \bibfield  {author} {\bibinfo {author} {\bibfnamefont {R.}~\bibnamefont
  {S{\"u}sstrunk}}\ and\ \bibinfo {author} {\bibfnamefont {S.~D.}\ \bibnamefont
  {Huber}},\ }\href@noop {} {\bibfield  {journal} {\bibinfo  {journal}
  {Science}\ }\textbf {\bibinfo {volume} {349}},\ \bibinfo {pages} {47}
  (\bibinfo {year} {2015})}\BibitemShut {NoStop}%
\bibitem [{\citenamefont {Mousavi}\ \emph {et~al.}(2015)\citenamefont
  {Mousavi}, \citenamefont {Khanikaev},\ and\ \citenamefont
  {Wang}}]{Mousavi2015}%
  \BibitemOpen
  \bibfield  {author} {\bibinfo {author} {\bibfnamefont {S.~H.}\ \bibnamefont
  {Mousavi}}, \bibinfo {author} {\bibfnamefont {A.~B.}\ \bibnamefont
  {Khanikaev}}, \ and\ \bibinfo {author} {\bibfnamefont {Z.}~\bibnamefont
  {Wang}},\ }\href@noop {} {\bibfield  {journal} {\bibinfo  {journal} {Nature
  communications}\ }\textbf {\bibinfo {volume} {6}} (\bibinfo {year}
  {2015})}\BibitemShut {NoStop}%
\bibitem [{\citenamefont {Yang}\ \emph {et~al.}(2015)\citenamefont {Yang},
  \citenamefont {Gao}, \citenamefont {Shi}, \citenamefont {Lin}, \citenamefont
  {Gao}, \citenamefont {Chong},\ and\ \citenamefont {Zhang}}]{Yang2015}%
  \BibitemOpen
  \bibfield  {author} {\bibinfo {author} {\bibfnamefont {Z.}~\bibnamefont
  {Yang}}, \bibinfo {author} {\bibfnamefont {F.}~\bibnamefont {Gao}}, \bibinfo
  {author} {\bibfnamefont {X.}~\bibnamefont {Shi}}, \bibinfo {author}
  {\bibfnamefont {X.}~\bibnamefont {Lin}}, \bibinfo {author} {\bibfnamefont
  {Z.}~\bibnamefont {Gao}}, \bibinfo {author} {\bibfnamefont {Y.}~\bibnamefont
  {Chong}}, \ and\ \bibinfo {author} {\bibfnamefont {B.}~\bibnamefont
  {Zhang}},\ }\href {\doibase 10.1103/PhysRevLett.114.114301} {\bibfield
  {journal} {\bibinfo  {journal} {Phys. Rev. Lett.}\ }\textbf {\bibinfo
  {volume} {114}},\ \bibinfo {pages} {114301} (\bibinfo {year}
  {2015})}\BibitemShut {NoStop}%
\bibitem [{\citenamefont {Peano}\ \emph {et~al.}(2015)\citenamefont {Peano},
  \citenamefont {Brendel}, \citenamefont {Schmidt},\ and\ \citenamefont
  {Marquardt}}]{Peano2015}%
  \BibitemOpen
  \bibfield  {author} {\bibinfo {author} {\bibfnamefont {V.}~\bibnamefont
  {Peano}}, \bibinfo {author} {\bibfnamefont {C.}~\bibnamefont {Brendel}},
  \bibinfo {author} {\bibfnamefont {M.}~\bibnamefont {Schmidt}}, \ and\
  \bibinfo {author} {\bibfnamefont {F.}~\bibnamefont {Marquardt}},\ }\href
  {\doibase 10.1103/PhysRevX.5.031011} {\bibfield  {journal} {\bibinfo
  {journal} {Phys. Rev. X}\ }\textbf {\bibinfo {volume} {5}},\ \bibinfo {pages}
  {031011} (\bibinfo {year} {2015})}\BibitemShut {NoStop}%
\bibitem [{\citenamefont {Strohm}\ \emph {et~al.}(2005)\citenamefont {Strohm},
  \citenamefont {Rikken},\ and\ \citenamefont {Wyder}}]{Strohm2005}%
  \BibitemOpen
  \bibfield  {author} {\bibinfo {author} {\bibfnamefont {C.}~\bibnamefont
  {Strohm}}, \bibinfo {author} {\bibfnamefont {G.}~\bibnamefont {Rikken}}, \
  and\ \bibinfo {author} {\bibfnamefont {P.}~\bibnamefont {Wyder}},\
  }\href@noop {} {\bibfield  {journal} {\bibinfo  {journal} {Phys. Rev. Lett.}\
  }\textbf {\bibinfo {volume} {95}},\ \bibinfo {pages} {155901} (\bibinfo
  {year} {2005})}\BibitemShut {NoStop}%
\bibitem [{\citenamefont {Sheng}\ \emph {et~al.}(2006)\citenamefont {Sheng},
  \citenamefont {Sheng},\ and\ \citenamefont {Ting}}]{Sheng2006}%
  \BibitemOpen
  \bibfield  {author} {\bibinfo {author} {\bibfnamefont {L.}~\bibnamefont
  {Sheng}}, \bibinfo {author} {\bibfnamefont {D.}~\bibnamefont {Sheng}}, \ and\
  \bibinfo {author} {\bibfnamefont {C.}~\bibnamefont {Ting}},\ }\href@noop {}
  {\bibfield  {journal} {\bibinfo  {journal} {Physical review letters}\
  }\textbf {\bibinfo {volume} {96}},\ \bibinfo {pages} {155901} (\bibinfo
  {year} {2006})}\BibitemShut {NoStop}%
\bibitem [{\citenamefont {Pal}\ \emph {et~al.}(2016)\citenamefont {Pal},
  \citenamefont {Schaeffer},\ and\ \citenamefont {Ruzzene}}]{Pal2016}%
  \BibitemOpen
  \bibfield  {author} {\bibinfo {author} {\bibfnamefont {R.~K.}\ \bibnamefont
  {Pal}}, \bibinfo {author} {\bibfnamefont {M.}~\bibnamefont {Schaeffer}}, \
  and\ \bibinfo {author} {\bibfnamefont {M.}~\bibnamefont {Ruzzene}},\
  }\href@noop {} {\bibfield  {journal} {\bibinfo  {journal} {Journal of Applied
  Physics}\ }\textbf {\bibinfo {volume} {119}},\ \bibinfo {pages} {084305}
  (\bibinfo {year} {2016})}\BibitemShut {NoStop}%
\bibitem [{\citenamefont {He}\ \emph {et~al.}(2016)\citenamefont {He},
  \citenamefont {Ni}, \citenamefont {Ge}, \citenamefont {Sun}, \citenamefont
  {Chen}, \citenamefont {Lu}, \citenamefont {Liu},\ and\ \citenamefont
  {Chen}}]{He2016}%
  \BibitemOpen
  \bibfield  {author} {\bibinfo {author} {\bibfnamefont {C.}~\bibnamefont
  {He}}, \bibinfo {author} {\bibfnamefont {X.}~\bibnamefont {Ni}}, \bibinfo
  {author} {\bibfnamefont {H.}~\bibnamefont {Ge}}, \bibinfo {author}
  {\bibfnamefont {X.-C.}\ \bibnamefont {Sun}}, \bibinfo {author} {\bibfnamefont
  {Y.-B.}\ \bibnamefont {Chen}}, \bibinfo {author} {\bibfnamefont {M.-H.}\
  \bibnamefont {Lu}}, \bibinfo {author} {\bibfnamefont {X.-P.}\ \bibnamefont
  {Liu}}, \ and\ \bibinfo {author} {\bibfnamefont {Y.-F.}\ \bibnamefont
  {Chen}},\ }\href@noop {} {\bibfield  {journal} {\bibinfo  {journal} {Nature
  Physics}\ } (\bibinfo {year} {2016})}\BibitemShut {NoStop}%
\bibitem [{\citenamefont {S{\"u}sstrunk}\ and\ \citenamefont
  {Huber}(2016)}]{Suesstrunk2016}%
  \BibitemOpen
  \bibfield  {author} {\bibinfo {author} {\bibfnamefont {R.}~\bibnamefont
  {S{\"u}sstrunk}}\ and\ \bibinfo {author} {\bibfnamefont {S.~D.}\ \bibnamefont
  {Huber}},\ }\href@noop {} {\bibfield  {journal} {\bibinfo  {journal} {PNAS}\
  }\textbf {\bibinfo {volume} {113}},\ \bibinfo {pages} {E4767} (\bibinfo
  {year} {2016})}\BibitemShut {NoStop}%
\bibitem [{\citenamefont {Xiao}\ \emph {et~al.}(2015)\citenamefont {Xiao},
  \citenamefont {Chen}, \citenamefont {He},\ and\ \citenamefont
  {Chan}}]{Xiao2015a}%
  \BibitemOpen
  \bibfield  {author} {\bibinfo {author} {\bibfnamefont {M.}~\bibnamefont
  {Xiao}}, \bibinfo {author} {\bibfnamefont {W.-J.}\ \bibnamefont {Chen}},
  \bibinfo {author} {\bibfnamefont {W.-Y.}\ \bibnamefont {He}}, \ and\ \bibinfo
  {author} {\bibfnamefont {C.}~\bibnamefont {Chan}},\ }\href@noop {} {\bibfield
   {journal} {\bibinfo  {journal} {Nature Physics}\ } (\bibinfo {year}
  {2015})}\BibitemShut {NoStop}%
\bibitem [{\citenamefont {Rocklin}\ \emph {et~al.}(2016)\citenamefont
  {Rocklin}, \citenamefont {Chen}, \citenamefont {Falk}, \citenamefont
  {Vitelli},\ and\ \citenamefont {Lubensky}}]{Rocklin2016}%
  \BibitemOpen
  \bibfield  {author} {\bibinfo {author} {\bibfnamefont {D.~Z.}\ \bibnamefont
  {Rocklin}}, \bibinfo {author} {\bibfnamefont {B.~G.~G.}\ \bibnamefont
  {Chen}}, \bibinfo {author} {\bibfnamefont {M.}~\bibnamefont {Falk}}, \bibinfo
  {author} {\bibfnamefont {V.}~\bibnamefont {Vitelli}}, \ and\ \bibinfo
  {author} {\bibfnamefont {T.~C.}\ \bibnamefont {Lubensky}},\ }\href {\doibase
  10.1103/PhysRevLett.116.135503} {\bibfield  {journal} {\bibinfo  {journal}
  {Physical Review Letters}\ }\textbf {\bibinfo {volume} {116}},\ \bibinfo
  {pages} {135503} (\bibinfo {year} {2016})}\BibitemShut {NoStop}%
\bibitem [{\citenamefont {Rocklin}\ \emph {et~al.}(2017)\citenamefont
  {Rocklin}, \citenamefont {Zhou}, \citenamefont {Sun},\ and\ \citenamefont
  {Mao}}]{Rocklin2017}%
  \BibitemOpen
  \bibfield  {author} {\bibinfo {author} {\bibfnamefont {D.~Z.}\ \bibnamefont
  {Rocklin}}, \bibinfo {author} {\bibfnamefont {S.}~\bibnamefont {Zhou}},
  \bibinfo {author} {\bibfnamefont {K.}~\bibnamefont {Sun}}, \ and\ \bibinfo
  {author} {\bibfnamefont {X.}~\bibnamefont {Mao}},\ }\href@noop {} {\bibfield
  {journal} {\bibinfo  {journal} {Nature Communications}\ }\textbf {\bibinfo
  {volume} {8}},\ \bibinfo {pages} {14201} (\bibinfo {year}
  {2017})}\BibitemShut {NoStop}%
\bibitem [{\citenamefont {Paulose}\ \emph
  {et~al.}(2015{\natexlab{a}})\citenamefont {Paulose}, \citenamefont
  {Meeussen},\ and\ \citenamefont {Vitelli}}]{Paulose2015a}%
  \BibitemOpen
  \bibfield  {author} {\bibinfo {author} {\bibfnamefont {J.}~\bibnamefont
  {Paulose}}, \bibinfo {author} {\bibfnamefont {A.~S.}\ \bibnamefont
  {Meeussen}}, \ and\ \bibinfo {author} {\bibfnamefont {V.}~\bibnamefont
  {Vitelli}},\ }\href@noop {} {\bibfield  {journal} {\bibinfo  {journal}
  {PNAS}\ }\textbf {\bibinfo {volume} {112}},\ \bibinfo {pages} {7639}
  (\bibinfo {year} {2015}{\natexlab{a}})}\BibitemShut {NoStop}%
\bibitem [{\citenamefont {Paulose}\ \emph
  {et~al.}(2015{\natexlab{b}})\citenamefont {Paulose}, \citenamefont {Chen},\
  and\ \citenamefont {Vitelli}}]{Paulose2015}%
  \BibitemOpen
  \bibfield  {author} {\bibinfo {author} {\bibfnamefont {J.}~\bibnamefont
  {Paulose}}, \bibinfo {author} {\bibfnamefont {B.~G.-g.}\ \bibnamefont
  {Chen}}, \ and\ \bibinfo {author} {\bibfnamefont {V.}~\bibnamefont
  {Vitelli}},\ }\href@noop {} {\bibfield  {journal} {\bibinfo  {journal}
  {Nature Physics}\ } (\bibinfo {year} {2015}{\natexlab{b}})}\BibitemShut
  {NoStop}%
\bibitem [{\citenamefont {Chen}\ \emph {et~al.}(2014)\citenamefont {Chen},
  \citenamefont {Upadhyaya},\ and\ \citenamefont {Vitelli}}]{Chen2014}%
  \BibitemOpen
  \bibfield  {author} {\bibinfo {author} {\bibfnamefont {B.~G.-g.}\
  \bibnamefont {Chen}}, \bibinfo {author} {\bibfnamefont {N.}~\bibnamefont
  {Upadhyaya}}, \ and\ \bibinfo {author} {\bibfnamefont {V.}~\bibnamefont
  {Vitelli}},\ }\href@noop {} {\bibfield  {journal} {\bibinfo  {journal}
  {Proceedings of the National Academy of Sciences}\ }\textbf {\bibinfo
  {volume} {111}},\ \bibinfo {pages} {13004} (\bibinfo {year}
  {2014})}\BibitemShut {NoStop}%
\bibitem [{\citenamefont {Souslov}\ \emph {et~al.}(2009)\citenamefont
  {Souslov}, \citenamefont {Liu},\ and\ \citenamefont
  {Lubensky}}]{Souslov2009}%
  \BibitemOpen
  \bibfield  {author} {\bibinfo {author} {\bibfnamefont {A.}~\bibnamefont
  {Souslov}}, \bibinfo {author} {\bibfnamefont {A.~J.}\ \bibnamefont {Liu}}, \
  and\ \bibinfo {author} {\bibfnamefont {T.~C.}\ \bibnamefont {Lubensky}},\
  }\href {\doibase 10.1103/PhysRevLett.103.205503} {\bibfield  {journal}
  {\bibinfo  {journal} {Phys. Rev. Lett.}\ }\textbf {\bibinfo {volume} {103}},\
  \bibinfo {eid} {205503} (\bibinfo {year} {2009})}\BibitemShut {NoStop}%
\bibitem [{\citenamefont {Mao}\ \emph {et~al.}(2010)\citenamefont {Mao},
  \citenamefont {Xu},\ and\ \citenamefont {Lubensky}}]{Mao2010}%
  \BibitemOpen
  \bibfield  {author} {\bibinfo {author} {\bibfnamefont {X.}~\bibnamefont
  {Mao}}, \bibinfo {author} {\bibfnamefont {N.}~\bibnamefont {Xu}}, \ and\
  \bibinfo {author} {\bibfnamefont {T.~C.}\ \bibnamefont {Lubensky}},\ }\href
  {\doibase 10.1103/PhysRevLett.104.085504} {\bibfield  {journal} {\bibinfo
  {journal} {Phys. Rev. Lett.}\ }\textbf {\bibinfo {volume} {104}},\ \bibinfo
  {pages} {085504} (\bibinfo {year} {2010})}\BibitemShut {NoStop}%
\bibitem [{\citenamefont {Ellenbroek}\ and\ \citenamefont
  {Mao}(2011)}]{Ellenbroek2011}%
  \BibitemOpen
  \bibfield  {author} {\bibinfo {author} {\bibfnamefont {W.~G.}\ \bibnamefont
  {Ellenbroek}}\ and\ \bibinfo {author} {\bibfnamefont {X.}~\bibnamefont
  {Mao}},\ }\href@noop {} {\bibfield  {journal} {\bibinfo  {journal} {Europhys.
  Lett.}\ }\textbf {\bibinfo {volume} {96}} (\bibinfo {year}
  {2011})}\BibitemShut {NoStop}%
\bibitem [{\citenamefont {Mao}\ and\ \citenamefont
  {Lubensky}(2011)}]{Mao2011a}%
  \BibitemOpen
  \bibfield  {author} {\bibinfo {author} {\bibfnamefont {X.}~\bibnamefont
  {Mao}}\ and\ \bibinfo {author} {\bibfnamefont {T.~C.}\ \bibnamefont
  {Lubensky}},\ }\href {\doibase 10.1103/PhysRevE.83.011111} {\bibfield
  {journal} {\bibinfo  {journal} {Phys. Rev. E}\ }\textbf {\bibinfo {volume}
  {83}},\ \bibinfo {pages} {011111} (\bibinfo {year} {2011})}\BibitemShut
  {NoStop}%
\bibitem [{\citenamefont {Sun}\ \emph {et~al.}(2012)\citenamefont {Sun},
  \citenamefont {Souslov}, \citenamefont {Mao},\ and\ \citenamefont
  {Lubensky}}]{Sun2012}%
  \BibitemOpen
  \bibfield  {author} {\bibinfo {author} {\bibfnamefont {K.}~\bibnamefont
  {Sun}}, \bibinfo {author} {\bibfnamefont {A.}~\bibnamefont {Souslov}},
  \bibinfo {author} {\bibfnamefont {X.}~\bibnamefont {Mao}}, \ and\ \bibinfo
  {author} {\bibfnamefont {T.~C.}\ \bibnamefont {Lubensky}},\ }\href@noop {}
  {\bibfield  {journal} {\bibinfo  {journal} {Proc. Natl. Acad. Sci. U. S. A.}\
  }\textbf {\bibinfo {volume} {109}},\ \bibinfo {pages} {12369} (\bibinfo
  {year} {2012})}\BibitemShut {NoStop}%
\bibitem [{\citenamefont {Zhang}\ \emph {et~al.}(2015)\citenamefont {Zhang},
  \citenamefont {Rocklin}, \citenamefont {Chen},\ and\ \citenamefont
  {Mao}}]{Zhang2015a}%
  \BibitemOpen
  \bibfield  {author} {\bibinfo {author} {\bibfnamefont {L.}~\bibnamefont
  {Zhang}}, \bibinfo {author} {\bibfnamefont {D.~Z.}\ \bibnamefont {Rocklin}},
  \bibinfo {author} {\bibfnamefont {B.~G.-g.}\ \bibnamefont {Chen}}, \ and\
  \bibinfo {author} {\bibfnamefont {X.}~\bibnamefont {Mao}},\ }\href {\doibase
  10.1103/PhysRevE.91.032124} {\bibfield  {journal} {\bibinfo  {journal} {Phys.
  Rev. E}\ }\textbf {\bibinfo {volume} {91}},\ \bibinfo {pages} {032124}
  (\bibinfo {year} {2015})}\BibitemShut {NoStop}%
\bibitem [{\citenamefont {Mao}\ \emph {et~al.}(2015)\citenamefont {Mao},
  \citenamefont {Souslov}, \citenamefont {Mendoza},\ and\ \citenamefont
  {Lubensky}}]{Mao2015}%
  \BibitemOpen
  \bibfield  {author} {\bibinfo {author} {\bibfnamefont {X.}~\bibnamefont
  {Mao}}, \bibinfo {author} {\bibfnamefont {A.}~\bibnamefont {Souslov}},
  \bibinfo {author} {\bibfnamefont {C.~I.}\ \bibnamefont {Mendoza}}, \ and\
  \bibinfo {author} {\bibfnamefont {T.~C.}\ \bibnamefont {Lubensky}},\
  }\href@noop {} {\bibfield  {journal} {\bibinfo  {journal} {Nature
  Communications}\ }\textbf {\bibinfo {volume} {6}},\ \bibinfo {pages} {5968}
  (\bibinfo {year} {2015})}\BibitemShut {NoStop}%
\bibitem [{\citenamefont {Sussman}\ \emph {et~al.}(2016)\citenamefont
  {Sussman}, \citenamefont {Stenull},\ and\ \citenamefont
  {Lubensky}}]{Sussman2016}%
  \BibitemOpen
  \bibfield  {author} {\bibinfo {author} {\bibfnamefont {D.~M.}\ \bibnamefont
  {Sussman}}, \bibinfo {author} {\bibfnamefont {O.}~\bibnamefont {Stenull}}, \
  and\ \bibinfo {author} {\bibfnamefont {T.}~\bibnamefont {Lubensky}},\
  }\href@noop {} {\bibfield  {journal} {\bibinfo  {journal} {Soft matter}\
  }\textbf {\bibinfo {volume} {12}},\ \bibinfo {pages} {6079} (\bibinfo {year}
  {2016})}\BibitemShut {NoStop}%
\bibitem [{\citenamefont {Mitchell}\ \emph {et~al.}(2016)\citenamefont
  {Mitchell}, \citenamefont {Nash}, \citenamefont {Hexner}, \citenamefont
  {Turner},\ and\ \citenamefont {Irvine}}]{Mitchell2016}%
  \BibitemOpen
  \bibfield  {author} {\bibinfo {author} {\bibfnamefont {N.~P.}\ \bibnamefont
  {Mitchell}}, \bibinfo {author} {\bibfnamefont {L.~M.}\ \bibnamefont {Nash}},
  \bibinfo {author} {\bibfnamefont {D.}~\bibnamefont {Hexner}}, \bibinfo
  {author} {\bibfnamefont {A.}~\bibnamefont {Turner}}, \ and\ \bibinfo {author}
  {\bibfnamefont {W.}~\bibnamefont {Irvine}},\ }\href@noop {} {\bibfield
  {journal} {\bibinfo  {journal} {arXiv preprint arXiv:1612.09267}\ } (\bibinfo
  {year} {2016})}\BibitemShut {NoStop}%
\bibitem [{\citenamefont {Head}\ \emph {et~al.}(2003)\citenamefont {Head},
  \citenamefont {Levine},\ and\ \citenamefont {MacKintosh}}]{Head2003}%
  \BibitemOpen
  \bibfield  {author} {\bibinfo {author} {\bibfnamefont {D.~A.}\ \bibnamefont
  {Head}}, \bibinfo {author} {\bibfnamefont {A.~J.}\ \bibnamefont {Levine}}, \
  and\ \bibinfo {author} {\bibfnamefont {F.~C.}\ \bibnamefont {MacKintosh}},\
  }\href {\doibase 10.1103/PhysRevLett.91.108102} {\bibfield  {journal}
  {\bibinfo  {journal} {Phys. Rev. Lett.}\ }\textbf {\bibinfo {volume} {91}},\
  \bibinfo {pages} {108102} (\bibinfo {year} {2003})}\BibitemShut {NoStop}%
\bibitem [{\citenamefont {Wilhelm}\ and\ \citenamefont
  {Frey}(2003)}]{Wilhelm2003}%
  \BibitemOpen
  \bibfield  {author} {\bibinfo {author} {\bibfnamefont {J.}~\bibnamefont
  {Wilhelm}}\ and\ \bibinfo {author} {\bibfnamefont {E.}~\bibnamefont {Frey}},\
  }\href {\doibase 10.1103/PhysRevLett.91.108103} {\bibfield  {journal}
  {\bibinfo  {journal} {Phys. Rev. Lett.}\ }\textbf {\bibinfo {volume} {91}},\
  \bibinfo {pages} {108103} (\bibinfo {year} {2003})}\BibitemShut {NoStop}%
\bibitem [{\citenamefont {Gardel}\ \emph {et~al.}(2004)\citenamefont {Gardel},
  \citenamefont {Shin}, \citenamefont {MacKintosh}, \citenamefont {Mahadevan},
  \citenamefont {Matsudaira},\ and\ \citenamefont {Weitz}}]{Gardel2004}%
  \BibitemOpen
  \bibfield  {author} {\bibinfo {author} {\bibfnamefont {M.}~\bibnamefont
  {Gardel}}, \bibinfo {author} {\bibfnamefont {J.}~\bibnamefont {Shin}},
  \bibinfo {author} {\bibfnamefont {F.}~\bibnamefont {MacKintosh}}, \bibinfo
  {author} {\bibfnamefont {L.}~\bibnamefont {Mahadevan}}, \bibinfo {author}
  {\bibfnamefont {P.}~\bibnamefont {Matsudaira}}, \ and\ \bibinfo {author}
  {\bibfnamefont {D.}~\bibnamefont {Weitz}},\ }\href@noop {} {\bibfield
  {journal} {\bibinfo  {journal} {Science}\ }\textbf {\bibinfo {volume}
  {304}},\ \bibinfo {pages} {1301} (\bibinfo {year} {2004})}\BibitemShut
  {NoStop}%
\bibitem [{\citenamefont {Storm}\ \emph {et~al.}(2005)\citenamefont {Storm},
  \citenamefont {Pastore}, \citenamefont {MacKintosh}, \citenamefont
  {Lubensky},\ and\ \citenamefont {Janmey}}]{Storm2005}%
  \BibitemOpen
  \bibfield  {author} {\bibinfo {author} {\bibfnamefont {C.}~\bibnamefont
  {Storm}}, \bibinfo {author} {\bibfnamefont {J.}~\bibnamefont {Pastore}},
  \bibinfo {author} {\bibfnamefont {F.}~\bibnamefont {MacKintosh}}, \bibinfo
  {author} {\bibfnamefont {T.}~\bibnamefont {Lubensky}}, \ and\ \bibinfo
  {author} {\bibfnamefont {P.}~\bibnamefont {Janmey}},\ }\href@noop {}
  {\bibfield  {journal} {\bibinfo  {journal} {Nature}\ }\textbf {\bibinfo
  {volume} {435}},\ \bibinfo {pages} {191} (\bibinfo {year}
  {2005})}\BibitemShut {NoStop}%
\bibitem [{\citenamefont {Heussinger}\ and\ \citenamefont
  {Frey}(2006)}]{Heussinger2006}%
  \BibitemOpen
  \bibfield  {author} {\bibinfo {author} {\bibfnamefont {C.}~\bibnamefont
  {Heussinger}}\ and\ \bibinfo {author} {\bibfnamefont {E.}~\bibnamefont
  {Frey}},\ }\href {\doibase 10.1103/PhysRevLett.97.105501} {\bibfield
  {journal} {\bibinfo  {journal} {Phys. Rev. Lett.}\ }\textbf {\bibinfo
  {volume} {97}},\ \bibinfo {pages} {105501} (\bibinfo {year}
  {2006})}\BibitemShut {NoStop}%
\bibitem [{\citenamefont {Broedersz}\ \emph {et~al.}(2011)\citenamefont
  {Broedersz}, \citenamefont {Mao}, \citenamefont {Lubensky},\ and\
  \citenamefont {MacKintosh}}]{Broedersz2011}%
  \BibitemOpen
  \bibfield  {author} {\bibinfo {author} {\bibfnamefont {C.~P.}\ \bibnamefont
  {Broedersz}}, \bibinfo {author} {\bibfnamefont {X.}~\bibnamefont {Mao}},
  \bibinfo {author} {\bibfnamefont {T.~C.}\ \bibnamefont {Lubensky}}, \ and\
  \bibinfo {author} {\bibfnamefont {F.~C.}\ \bibnamefont {MacKintosh}},\
  }\href@noop {} {\bibfield  {journal} {\bibinfo  {journal} {Nat. Phys.}\
  }\textbf {\bibinfo {volume} {7}},\ \bibinfo {pages} {983} (\bibinfo {year}
  {2011})}\BibitemShut {NoStop}%
\bibitem [{\citenamefont {Mao}\ \emph {et~al.}(2013{\natexlab{a}})\citenamefont
  {Mao}, \citenamefont {Stenull},\ and\ \citenamefont {Lubensky}}]{Mao2013b}%
  \BibitemOpen
  \bibfield  {author} {\bibinfo {author} {\bibfnamefont {X.}~\bibnamefont
  {Mao}}, \bibinfo {author} {\bibfnamefont {O.}~\bibnamefont {Stenull}}, \ and\
  \bibinfo {author} {\bibfnamefont {T.~C.}\ \bibnamefont {Lubensky}},\ }\href
  {\doibase 10.1103/PhysRevE.87.042601} {\bibfield  {journal} {\bibinfo
  {journal} {Phys. Rev. E}\ }\textbf {\bibinfo {volume} {87}},\ \bibinfo
  {pages} {042601} (\bibinfo {year} {2013}{\natexlab{a}})}\BibitemShut
  {NoStop}%
\bibitem [{\citenamefont {Mao}\ \emph {et~al.}(2013{\natexlab{b}})\citenamefont
  {Mao}, \citenamefont {Stenull},\ and\ \citenamefont {Lubensky}}]{Mao2013c}%
  \BibitemOpen
  \bibfield  {author} {\bibinfo {author} {\bibfnamefont {X.}~\bibnamefont
  {Mao}}, \bibinfo {author} {\bibfnamefont {O.}~\bibnamefont {Stenull}}, \ and\
  \bibinfo {author} {\bibfnamefont {T.~C.}\ \bibnamefont {Lubensky}},\ }\href
  {\doibase 10.1103/PhysRevE.87.042602} {\bibfield  {journal} {\bibinfo
  {journal} {Phys. Rev. E}\ }\textbf {\bibinfo {volume} {87}},\ \bibinfo
  {pages} {042602} (\bibinfo {year} {2013}{\natexlab{b}})}\BibitemShut
  {NoStop}%
\bibitem [{\citenamefont {Broedersz}\ and\ \citenamefont
  {MacKintosh}(2014)}]{Broedersz2014}%
  \BibitemOpen
  \bibfield  {author} {\bibinfo {author} {\bibfnamefont {C.~P.}\ \bibnamefont
  {Broedersz}}\ and\ \bibinfo {author} {\bibfnamefont {F.~C.}\ \bibnamefont
  {MacKintosh}},\ }\href@noop {} {\bibfield  {journal} {\bibinfo  {journal}
  {Reviews of Modern Physics}\ }\textbf {\bibinfo {volume} {86}},\ \bibinfo
  {pages} {995} (\bibinfo {year} {2014})}\BibitemShut {NoStop}%
\bibitem [{\citenamefont {Sharma}\ \emph {et~al.}(2016)\citenamefont {Sharma},
  \citenamefont {Licup}, \citenamefont {Jansen}, \citenamefont {Rens},
  \citenamefont {Sheinman}, \citenamefont {Koenderink},\ and\ \citenamefont
  {MacKintosh}}]{Sharma2016}%
  \BibitemOpen
  \bibfield  {author} {\bibinfo {author} {\bibfnamefont {A.}~\bibnamefont
  {Sharma}}, \bibinfo {author} {\bibfnamefont {A.}~\bibnamefont {Licup}},
  \bibinfo {author} {\bibfnamefont {K.}~\bibnamefont {Jansen}}, \bibinfo
  {author} {\bibfnamefont {R.}~\bibnamefont {Rens}}, \bibinfo {author}
  {\bibfnamefont {M.}~\bibnamefont {Sheinman}}, \bibinfo {author}
  {\bibfnamefont {G.}~\bibnamefont {Koenderink}}, \ and\ \bibinfo {author}
  {\bibfnamefont {F.}~\bibnamefont {MacKintosh}},\ }\href@noop {} {\bibfield
  {journal} {\bibinfo  {journal} {Nature Physics}\ }\textbf {\bibinfo {volume}
  {12}},\ \bibinfo {pages} {584} (\bibinfo {year} {2016})}\BibitemShut
  {NoStop}%
\bibitem [{\citenamefont {Feng}\ \emph {et~al.}(2015)\citenamefont {Feng},
  \citenamefont {Levine}, \citenamefont {Mao},\ and\ \citenamefont
  {Sander}}]{Feng2015}%
  \BibitemOpen
  \bibfield  {author} {\bibinfo {author} {\bibfnamefont {J.}~\bibnamefont
  {Feng}}, \bibinfo {author} {\bibfnamefont {H.}~\bibnamefont {Levine}},
  \bibinfo {author} {\bibfnamefont {X.}~\bibnamefont {Mao}}, \ and\ \bibinfo
  {author} {\bibfnamefont {L.~M.}\ \bibnamefont {Sander}},\ }\href@noop {}
  {\bibfield  {journal} {\bibinfo  {journal} {Physical Review E}\ }\textbf
  {\bibinfo {volume} {91}},\ \bibinfo {pages} {042710} (\bibinfo {year}
  {2015})}\BibitemShut {NoStop}%
\bibitem [{\citenamefont {Feng}\ \emph {et~al.}(2016)\citenamefont {Feng},
  \citenamefont {Levine}, \citenamefont {Mao},\ and\ \citenamefont
  {Sander}}]{Feng2016}%
  \BibitemOpen
  \bibfield  {author} {\bibinfo {author} {\bibfnamefont {J.}~\bibnamefont
  {Feng}}, \bibinfo {author} {\bibfnamefont {H.}~\bibnamefont {Levine}},
  \bibinfo {author} {\bibfnamefont {X.}~\bibnamefont {Mao}}, \ and\ \bibinfo
  {author} {\bibfnamefont {L.~M.}\ \bibnamefont {Sander}},\ }\href@noop {}
  {\bibfield  {journal} {\bibinfo  {journal} {Soft matter}\ }\textbf {\bibinfo
  {volume} {12}},\ \bibinfo {pages} {1419} (\bibinfo {year}
  {2016})}\BibitemShut {NoStop}%
\bibitem [{\citenamefont {Alberts}\ \emph {et~al.}(2008)\citenamefont
  {Alberts}, \citenamefont {Johnson}, \citenamefont {Lewis}, \citenamefont
  {Raff}, \citenamefont {Roberts},\ and\ \citenamefont {Walter}}]{Alberts2008}%
  \BibitemOpen
  \bibfield  {author} {\bibinfo {author} {\bibfnamefont {B.}~\bibnamefont
  {Alberts}}, \bibinfo {author} {\bibfnamefont {A.}~\bibnamefont {Johnson}},
  \bibinfo {author} {\bibfnamefont {J.}~\bibnamefont {Lewis}}, \bibinfo
  {author} {\bibfnamefont {M.}~\bibnamefont {Raff}}, \bibinfo {author}
  {\bibfnamefont {K.}~\bibnamefont {Roberts}}, \ and\ \bibinfo {author}
  {\bibfnamefont {P.}~\bibnamefont {Walter}},\ }\href@noop {} {\emph {\bibinfo
  {title} {Molecular Biology of the Cell}}},\ \bibinfo {edition} {4th}\ ed.\
  (\bibinfo  {publisher} {Garland, New York},\ \bibinfo {year}
  {2008})\BibitemShut {NoStop}%
\bibitem [{\citenamefont {Brangwynne}\ \emph {et~al.}(2008)\citenamefont
  {Brangwynne}, \citenamefont {Koenderink}, \citenamefont {MacKintosh},\ and\
  \citenamefont {Weitz}}]{Brangwynne2008}%
  \BibitemOpen
  \bibfield  {author} {\bibinfo {author} {\bibfnamefont {C.~P.}\ \bibnamefont
  {Brangwynne}}, \bibinfo {author} {\bibfnamefont {G.~H.}\ \bibnamefont
  {Koenderink}}, \bibinfo {author} {\bibfnamefont {F.~C.}\ \bibnamefont
  {MacKintosh}}, \ and\ \bibinfo {author} {\bibfnamefont {D.~A.}\ \bibnamefont
  {Weitz}},\ }\href@noop {} {\bibfield  {journal} {\bibinfo  {journal} {The
  Journal of cell biology}\ }\textbf {\bibinfo {volume} {183}},\ \bibinfo
  {pages} {583} (\bibinfo {year} {2008})}\BibitemShut {NoStop}%
\bibitem [{\citenamefont {Joanny}\ and\ \citenamefont
  {Prost}(2009)}]{Joanny2009}%
  \BibitemOpen
  \bibfield  {author} {\bibinfo {author} {\bibfnamefont {J.-F.}\ \bibnamefont
  {Joanny}}\ and\ \bibinfo {author} {\bibfnamefont {J.}~\bibnamefont {Prost}},\
  }\href@noop {} {\bibfield  {journal} {\bibinfo  {journal} {HFSP journal}\
  }\textbf {\bibinfo {volume} {3}},\ \bibinfo {pages} {94} (\bibinfo {year}
  {2009})}\BibitemShut {NoStop}%
\end{thebibliography}
\end{document}